\documentclass[amsmath,amssymb,nofootinbib,showpacs,twocolumn,nofootinbib]{revtex4-1}
\usepackage{dcolumn}
\usepackage{bm}
\usepackage{graphicx}
\usepackage{epsfig}
\usepackage{color}
\usepackage{natbib}
\usepackage{twoopt}
\usepackage{dashrule}

\usepackage{amsmath}
\usepackage{wasysym}
\usepackage{xspace}
\usepackage[dvipsnames]{xcolor}

\usepackage[breaklinks=true,colorlinks=true,linkcolor=RubineRed,citecolor=RubineRed,urlcolor=RubineRed,pdfauthor={Tomassetti},pdftitle={Stochastic}]{hyperref}
\def\MyTitle#1{{\noindent\textit{#1}}\,\,---\,} 

\newcommand{\GCR}{CR\xspace}
\newcommand{\GCRs}{CRs\xspace}
\newcommand{\etal}{et al.}
\newcommand{\eg}{\textit{e.g.}} 
\newcommand{\ie}{\textit{i.e.}} 

\newcommand{\ApJ}{Astrophys. J.}
\newcommand{\AeA}{Astron. \& Astrophys.}
\newcommand{\PRL}{Phys. Rev. Lett.}
\newcommand{\PRD}{Phys. Rev. D}
\newcommand{\ASR}{Adv. Space Res.}

\newcommand{\MNRAS}{MNRAS}
\newcommand{\ApSS}{Astrophys. Space Sci.}
\newcommand{\GRL}{Geophys. Res. Lett.}
\newcommand{\JGR}{J. Geophys. Res.}
\newcommand{\SolPhys}{Solar Phys.}

\def\citem#1{Ref.\,\cite{#1}}

\begin{document}
%
\title{Numerical modeling of cosmic rays in the heliosphere: \\Analysis of proton data from AMS-02 and PAMELA}
\author{E. Fiandrini$^{\,1}$,  N. Tomassetti$^{\,1}$, B. Bertucci$^{\,1}$, F. Donnini$^{\,2}$, M. Graziani$^{\,1}$, B. Khiali$^{\,3}$, A. Reina Conde$^{\,4}$}
\address{$^{1}$\,Dipartimento di Fisica e Geologia, University of Perugia, Italy}
\address{$^{2}$\,INFN - Sezione di Perugia, Italy}
\address{$^{3}$\,INFN - Sezione di Roma Tor Vergata \& ASI Space Science Data Center (SSDC), Roma, Italy}
\address{$^{4}$\,Instituto de Astrofísica de Canarias (IAC), Universidad de La Laguna, Tenerife, Spain}

\begin{abstract}  
  Galactic cosmic rays (\GCRs) inside the heliosphere are affected by solar modulation.
  To investigate this phenomenon and its underlying physical mechanisms, we have performed
  a data-driven analysis of the temporal dependence of the \GCR proton flux over the solar cycle.
  The modulation effect was modeled by means of stochastic simulations of cosmic particles in the heliosphere.
  The model were constrained using measurements of \GCR protons made by AMS-02 and PAMELA experiments on monthly basis from 2006 to 2017.  
  With a global statistical analysis of these data, we have determined the key model parameters governing \GCR diffusion,
  its dependence on the particle rigidity, and its evolution over the solar cycle.
  Our results span over epochs of solar minimum, solar maximum, as well as epochs with magnetic reversal and opposite polarities.
  Along with the evolution of the \GCR transport parameters, we study their relationship with solar activity proxies and interplanetary parameters.
  We find that the rigidity dependence of the parallel mean free path of \GCR diffusion shows a remarkable time dependence,
  indicating a long-term variability in the interplanetary turbulence that interchanges across different regimes over the solar cycle.
  The evolution of the diffusion parameters show a delayed correlation with solar activity proxies, 
  reflecting the dynamics of the heliospheric plasma, and distinct dependencies for opposite states of magnetic polarity,
  reflecting the influence of charge-sign dependent drift in the \GCR modulation.
\end{abstract}  

\maketitle

\section{Introduction}     
\label{Sec::Introduction}  

Galactic  cosmic  rays  (\GCR)  are  high-energy charged  particles produced by astrophysical sources, distributed in our galaxy,
which travel through the interstellar medium and finally arrive at the
boundary of the nearby region to Earth where the Sun's activity dominates: the so called {\it heliosphere}.
When entering the heliosphere, \GCRs travel against the expanding solar wind (SW) and 
interact with the turbulent heliospheric magnetic field (HMF) \citep{Potgieter2013}. 
They are subjected to basic transport processes such as convection, diffusion and adiabatic energy losses. 
They are also subjected to the gradient-curvature drifts in the large-scale HMF and to the effects of the heliospheric current sheet (HCS).
Magnetic drift depends on the charge-sign of the particles and on the polarity of the HMF; \GCRs drift along different trajectories according to the polarity of the HMF. 
The cumulative effects of these processes are behind the so-called \emph{solar modulation} phenomenon of \GCRs, that is,
the modification of the energy spectra of \GCRs in the heliosphere, which is driven by the Sun's magnetic activity. 
Due to solar modulation, the \GCR flux observed at Earth is significantly different from that in interstellar space, known as Local Interstellar Spectrum (LIS).
Solar modulation depends on the \GCR particle species, its energy, and its charge sign.
It is also a time-dependent and space-dependent phenomenon, \ie, it depends on where and when the \GCR flux is measured inside the heliosphere.
The solar modulation effect decreases with increasing energy of the \GCR particles.
With the precision of the new \GCR data from AMS-02, the modulation effect is appreciable at kinetic energies up to dozens GeV. 
Solar activity shows a 11-year cycle, from its minimum when the Sun is quiet and the \GCR intensity is at its largest,
to its maximum of solar activity when the \GCR flux is minimum.
The intensity and the energy spectra of the \GCR flux are therefore anti-correlated with solar activity, in relation with its varying proxies such as the
number of sunspot (SSN) or the tilt angle of the solar magnetic axis with respect to the rotation axis $\alpha$ \citep{Usoskin1998,RossChaplin2019,Hoeksema1995}.
Along with the 11-year solar cycle, the HMF polarity shows a remarkable 22-year periodicity, with the magnetic reversal occurring during each maximum of solar activity. 
This periodicity is important for \GCR modulation, and in particular to study the effects of particle drifts in the large-scale HMF.

Since \GCR modulation is a manifestation of the \GCR propagation through the heliosphere,
\GCR data can be used to investigate the fundamental physics processes governing the transport of charged particles through the heliospheric plasma. 
In particular, precise measurements of the energy and time dependence of the \GCR fluxes may help to disentangle the interplay of
the different physics mechanisms at work. 
In this respect, the physical understanding of \GCR modulation in the heliosphere is one of the main objectives of
many theoretical and observational studies \citep{Corti2019,Boschini2017,Bobik2016,Potgieter2017}. 
Besides, modeling the \GCR modulation is essential for the search of new physics signatures in the fluxes of \GCR antimatter such as positrons or antiprotons.
An antimatter excess in \GCRs may suggest the occurrence of dark matter annihilation processes or the existence of new astrophysical sources of antimatter. 
Since the low-energy spectra of \GCRs are influenced by solar modulation, any interpretation about the origin of antiparticles
requires an accurate modeling of the charge-sign and energy dependent effects of \GCR modulation \citep{Tomassetti2017BCUnc}. 
Understanding the evolution of the \GCR fluxes in the heliosphere is also important for assessing the radiation hazard
of astronauts, electronics, and communication systems for low-Earth-orbit satellites or deep space missions \citep{Norbury2018,Mrigakshi2012}. 
In fact, the Galactic \GCR flux constitutes a significant dose of ionizing radiation for human bodies and electronics, 
and thus an accurate knowledge of the temporal and spatial variation of the \GCR in the heliosphere will reduce the
uncertainties in the radiation dose evaluation \citep{Cucinotta2015}. 
An important challenge, in this context, is to establish a predictive model for solar modulation that
is able to forecast the \GCR flux evolution using solar activity proxies.

From the observational point of view, a substantial progress has been made with the new measurements of the proton flux
from the Alpha Magnetic Spectrometer (AMS-02) experiment in the International Space Station \citep{Aguilar2018PHeVSTime, Aguilar2018LeptonsVSTime}
and the PAMELA mission onboard the Resurs-DK1 satellite \citep{Adriani2013Protons,Martucci2018},
along with the data provided by the Voyager-1 spacecraft beyond the heliosphere \citep{Cummings2016}.
In particular, AMS-02 and PAMELA have recently released accurate measurements of \GCR proton spectra over Bartels' rotation basis (BR, 27 days),
over extended energy range and for extended time periods, covering the  long solar minimum of 2006-2009 (cycle 23/24), the ascending phase of cycle 24,
the solar maximum and HMF reversal of 2013-2014, and the subsequent descending phase towards the new minimum until May 2017. 
Therefore, the data allows for the study of the \GCR propagation in the heliosphere under very different conditions of solar activity 
and epochs of opposite HMF polarities, which may bring a substantial advance in the understanding of the solar modulation phenomenon.

In this paper, we present a data-driven analysis of the temporal dependence of the flux of \GCR protons,
which constitute the most abundant species of the Galactic cosmic radiation.
The analysis has been conducted using a stochastic model of \GCR propagation,
\ie, a Monte Carlo based approach in which the solar modulation effect is computed by statistical sampling. 
Using the recent time- and energy-resolved measurements of \GCR proton fluxes on BR basis,
by means of a procedure of statistical inference, we determine the temporal and rigidity dependencies of the
mean free path of \GCRs propagating through the heliosphere, along with the corresponding uncertainties.
The rest of this paper is organized as follows. In Sect.\,\ref{Sec::Model}, we describe in details 
the numerical implementation of the \GCR modulation model, which is based on known and conventional mechanisms of particle transport in the heliosphere.
In Sect.\,\ref{Sec::Fitting} we present the procedure for the data-driven determination of the key model parameters and their uncertainty,
which is based on a grid sampling over a multidimensional parameter space.
In Sect.\,\ref{Sec::Results} we present the fit results and discuss their interpretation, in terms of physical mechanisms of \GCR transport,
in relation with the properties of heliospheric environment or with known proxies of solar activity.
We then conclude, in Sect.\,\ref{Sec::Conclusions}, with a summary of our study and a discussion on its future developments.

\section{The numerical model}  
\label{Sec::Model}             

To get a realistic description of \GCR modulation phenomenon, one needs to capture the essential features of \GCR transport in the heliosphere.
The diffusive propagation of the charged particles in the turbulent heliospheric plasma is described by the Parker's equation \citep{Parker1965}:
\begin{equation}
\label{eqn-park1}
\begin{split}
\frac{\partial f}{\partial t} + \nabla\cdot (\vec{V}_{\rm{sw}} - \mathbf{K}\cdot\nabla f) 
& - \frac{1}{3}(\nabla \cdot\vec{V}_{\rm{sw}})\frac{\partial f}{\partial (\ln\!{R})}  = 0 \,.
\end{split}
\end{equation}
The equation, along with its boundary conditions, describes the evolution of the distribution function $f(t,\vec{r},R)$ for a given particle species,
where {\it t} is the time, and $R$ is the particle rigidity, \ie,  the momentum per charge units $R{=}p/Z$. In this paper, we will focus on cosmic protons,
so that $R\equiv{p}$. The quantity $\mathbf{K}$ is the drift-diffusion tensor of the \GCR particles in the turbulent HMF of the heliosphere.

Because of the complexity of the transport equation, analytical solutions can be found only for very simplified situations such as in the
Force-Field or the Diffusion-Convection approximations \citep{Moraal2013,Zhu2018}. The full solution of Eq.(\ref{eqn-park1}) can be obtained numerically. 
Here we employ the stochastic method, that has become widely implemented in recent years thanks to the enormous progress in computing speed
and resources \citep{Potgieter2017,Kappl2016,Boschini2018}. 
The method consists of transforming the Parker's equation into a set of Stochastic Differential Equations (SDE) and then using Monte Carlo
simulations to sample the solution, \ie, the differential \GCR intensity for a given species, at a given position in heliosphere \citep{Strauss2017,Kopp2012}.

In general, the flux of \GCRs inside the heliosphere is time-dependent, reflecting the
varying conditions of the medium over which they propagate \citep{Ferreira2004}. 
A common practice is to follow a quasi steady-state approximation where
the time-dependent \GCR modulation is described as a succession of steady-state solutions ($\partial/\partial t=0$)
and the effective status of the heliospheric plasma during the \GCR propagation is defined in a suitable way.
The approximate way of taking into account the varying status of the heliosphere during the \GCR propagation is described in Sect.\,\ref{Sec::Model}.
Furthermore, in the SDE method, pseudo-particles are propagated backward in time from the Earth position to the heliospheric boundaries.
The numerical engine for handling the Monte Carlo generation and the trajectory tracing
is extracted from the publicly available code SolarProp \citep{Kappl2016}.
Based on the SolarProp simulation framework, we have implemented a customized model that is described in the following.

\subsection{The modulation region}  
\label{Sec::ModulationRegion}       

The heliosphere is a dynamic void in the ISM generated by the SW and regulated by Sun's activity.
The relevant boundary for the \GCR modulation phenomenon is the heliopause (HP), which separates the heliospheric plasma from the local ISM.
The HP is usually modeled as a spherical structure of radius $r_{\rm{HP}}\,\approx$\,122\,AU, where the Sun lies at its center.
Within the heliosphere, the termination shock (TS) is located at $r_{\rm{TS}}\cong$\,85\,AU, while the Earth position is
at $r_{0}\,\equiv$\,1\,AU placed in the equatorial plane.

\MyTitle{The large-scale HMF} 
%
The outward flowing SW embeds a frozen-in HMF which is wounded up in a modified Parker spiral \citep{Parker1958}. The ideal Parker's field is given by:
\begin{equation}
\label{parker_ideal}
\vec{B} = AB_0\left(\frac{r_{0}}{r}\right)^2\left(   \hat{e}_r - \tan\psi\hat{e}_{\phi}    \right)\left[  1 - 2H\left(  \theta - \Theta\right) \right] \,,
\end{equation}
where $r$ and $\theta$ are helioradius and colatitude, $B_0$ is the HMF value at Earth position, $A=\pm\,1$ is the field polarity, and $H$ is
the Heavyside step function.
The winding angle $\psi$ of the field line is defined as $\tan\psi = \Omega(r - r_{\odot})\sin\theta/V_{\rm{sw}}$;
the angle $\Theta$ determines the position of the wavy HCS, given
by $\Theta = \pi/2 + \sin^{-1}\left[\sin\alpha \sin \left(\Omega r/V_w\right)\right]$ \citep{JokipiiThomas1981}.
Here the quantity $\Omega$ is the average equatorial rotation
speed $\approx 2.73\times$$10^{-6}$ rad s$^{-1}$, $\alpha$ is the HCS tilt angle and $r_{\odot}$\,=\,696.000\,km is the radius of the Sun.
The Parker's model overwounds by several degrees beyond the value of the winding angle $\psi$, determined by the model at the polar regions.
%
\begin{figure}[b!]
\centering
\includegraphics[width=0.4\textwidth]{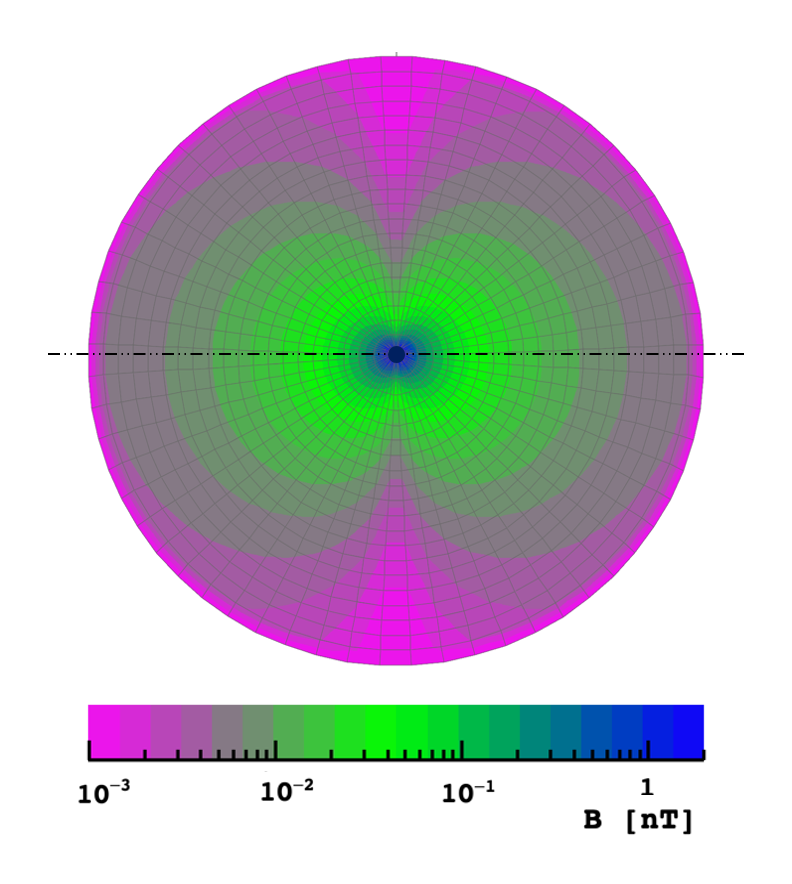}
\caption{Side view of the HMF field model in the $(x,z)$ plane of the heliosphere. The dashed line is the equatorial plane.}
\label{Fig::ccHMFModel}
\end{figure}
%
%
%
To avoid this, one has to consider that solar wind disturbances and plasma waves propagating along the open field lines
modify the magnetic field at the polar regions, so that it does not degenerate to a straight line along the polar axis.
Here we adopt the modification of \citet{JokipiiKota1989}:
\begin{equation}
\label{kota_corr}
B = B_0\left(\frac{r_0}{r}\right)^{2} \left\{  1 + \tan^2\psi + \left(\frac{r\delta(\theta)}{r_{\odot}} \right)^{2} \right\}^{1/2} \,,
\end{equation}
 where $\delta(\theta) = 8.7\times10^{-5}/\sin(\theta)$ if $1.7^{\circ} < \theta < 178.3^{\circ}$ and $\simeq$3$\times10^{-3}$ otherwise \citep{Fichtner1996}.
The winding angle $\psi$ is then modified as:
\begin{equation}
\label{kota_corr_psi}
\tan\psi = \left\{\frac{\Omega(r - r_{\odot})}{V}  + \left(\frac{r\delta(\theta)}{r_{\odot}} \right)^2 \right\}^{1/2}  \,.
\end{equation}
The term involving the dimensionless constant $\delta$ reflects the fact that the random field  is equivalent to a small latitudinal
component $B_{\theta} \sim\delta(\theta)r/r_{\odot}$. 
In this way, modifications on HMF and winding angle are effective only near the polar regions, as shown in Fig.~\ref{Fig::ccHMFModel}
where the two quantities are shown as function of colatitude.
It is worth noticing that the definitions of $B_{\theta}$ and $\delta(\theta)$ imply $\vec{\nabla}\cdot\vec{B} = 0$.

\MyTitle{Polarity and Tilt Angle} 
An important characteristic for the \GCR solar modulation is that the HMF follows a $\sim$\,22-year cycle,
known as magnetic polarity cycle, characterized by a N/S reversal about every $\sim$\,11 years, during the maximum of solar activity.
The period when $\vec{B}$ is directed outwards in the northern hemisphere of the Sun is known as positive polarity epoch($A>0$),
while when it has the opposite direction are known as ($A<0$) cycle.
In practice the quantity $A$ is a dichotomous variable that expresses the sign of $B$-field projection in the outward direction from the northern
hemisphere, $A\equiv{B_{N}/|B_{N}|}$ (or the inward projection of $B_{S}$ in the southern hemisphere).
In practice it can be determined using observations of the polar HMF in proximity of the Sun (Sect.\,\ref{Sec::TheParameters}).
The relevance of magnetic polarity in the context of solar modulation arises from \GCR drift motion:
it can be seen (Sect.\,\ref{Sec::Transport}) that the equations ruling \GCR drift in the HMF depend
upon the sign of the product between $A$ and $\hat{q}=Q/|Q|$, where $Q$ is the \GCR electric charge.
Thus, opposite drift directions are expected for opposite $\hat{q}A$ conditions.
A major co-rotating structure relevant to \GCR modulation is the HCS, which divides the HMF into hemispheres of opposite (N/S) polarity and where $B=0$. 
Due to the tilt of the solar magnetic axis, the HCS is wavy. 
The level of the HCS wavyness changes with time and it is set by the tilt angle $\alpha(t)$.
Typically, it varies from $\alpha\sim\,5^{\circ}$ during solar minimum to  $\alpha\sim\,70^{\circ}$ during solar maximum.
The tilt angle is reconstructed by the Wilcox Solar Observatory using two different models for the polar magnetic field: the so-called L-model and R-model.
In this work the classical L-model reconstruction is used as default.

\MyTitle{The Wind} 
%
The SW speed $V_{sw}$ is taken as radially directed outward.
However, the wind field exhibits a radial, latitudinal, and temporal dependence, where the latter is related to the solar cycle. 
During periods of solar minimum, the flow becomes distinctively latitude dependent, changing from $\sim$400 km\,s$^{-1}$
in the equatorial plane (slow speed region) to $\sim$\,800 km\,s$^{-1}$ in the polar regions (high speed region), as observed by Ulysses\,\citep{Ulysses1}. 
This effect is mitigated during epochs of solar maximum, when the angular extension of the slow-speed region increases to higher latitudes.
Beyond the TS, the SW slows down by a factor $1/S$, where $S=2.5$ is the shock compression ratio, as measured by the Voyager probes \citep{Cummings2005}. 
In this region, the wind is slowed down to subsonic speed.
%
\begin{figure}[t!]
\centering
\includegraphics[width=0.45\textwidth]{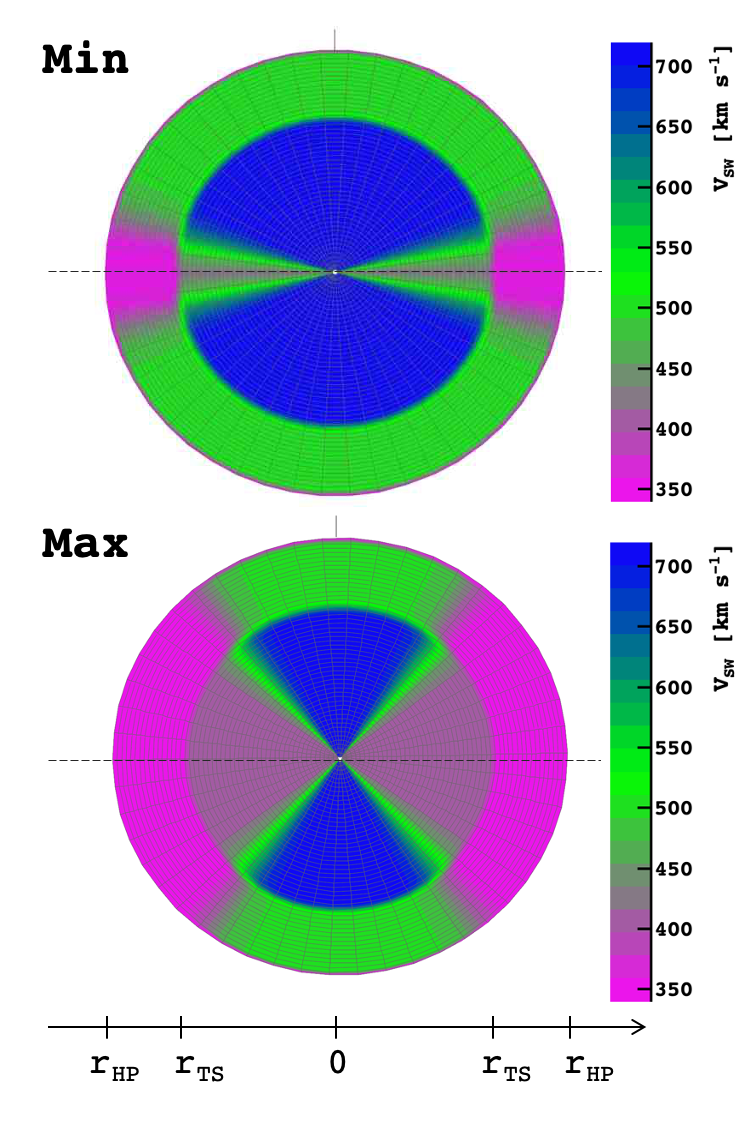}
\caption{Side view of the SW speed profile in the $(x,z)$ of the heliosphere, showing its latitudinal dependence in the typical cases of solar minimum
  (\texttt{Min}, for $\alpha{\cong}10^{\circ}$) and solar maximum (\texttt{Max}, for $\alpha{\cong}60^{\circ}$),
  where the latitudinal transition from a slow to a fast region depends on the HCS tilt angle $\alpha$.}
\label{swspeed}
\end{figure}
%
To incorporate such features in our model, we adopt the parametric expression given in \citep{Potgieter2014}:
\begin{equation}
\label{swprofile}
\begin{aligned}
V_{\rm{sw}}(r,\theta) = & V_0\left\{   1.475 \mp 0.4 \tanh\left[  6.8\left(  \theta -\pi/2 \pm \theta_T \right) \right]  \right\} \\
&  \times\left[    \frac{S+1}{2S} - \frac{S-1}{2S}\tanh\left(   \frac{r - r_{\rm{TS}}}{L}   \right)   \right] \,,
\end{aligned}
\end{equation} 
where $V_{0}$ = 400 km\,s$^{-1}$, and $L$ = 1.2 AU is the scale thickness of the TS.
The top and bottom signs correspond to the northern ($0\leq \theta \leq \pi/2$) and southern hemisphere ($\pi/2 \leq \theta \leq \pi$) of the heliosphere, respectively. 
The angle $\theta_{T}$ determines the polar angle at which the SW speed changes from a slow to a fast region. It is defined as $\theta_{T}= \alpha + \delta\alpha$,
where $\alpha$ is the tilt angle of the HCS and $\delta\alpha = 10^{\circ}$ is the width of the transition.
With this approach, the angular extension $\theta_{T}$ of the SW profile changes in time and it is linked to the level of solar activity, using the angle $\alpha$ as proxy.
The expression is valid for $r\gg{r_{\odot}}$, \ie, away from the Sun. Beyond the TS, the real SW speed is expected
to decrease as $r^{-2}$, so that $\vec{\nabla}\cdot\vec{V_{sw}}=0$ and \GCR particles do not experience adiabatic cooling.
The radial and latitudinal SW profile is shown in Fig.~\ref{swspeed} for two values of $\alpha$ corresponding
to solar minimum ($\alpha\cong\,10^{\circ}$) and solar maximum ($\alpha\cong\,60^{\circ}$) conditions.

\subsection{The particle transport}  
\label{Sec::Transport}               
%
%
The Parker's equation for the particle transport contains all physical processes
experienced by a given species of \GCR particles traveling in the interplanetary space.
In Eq.(\ref{eqn-park1}), the drift-diffusion tensor can be written as:
\begin{equation}
\label{diff-tens}
\mathbf{K} =
\begin{bmatrix}
K_{r\perp} &     -K_{A}                & 0  \\
K_{A}        &      K_{\theta\perp} & 0 \\
0               &        0                 & K_{\parallel}
\end{bmatrix} 
\end{equation}
in a reference system with the third coordinate along the average magnetic field.
The symbol $K_{\parallel}$ denotes the diffusion coefficient along the field direction,
while $K_{\theta\perp}$ and $K_{r\perp}$ the diffusion coefficients along the perpendicular and radial direction, respectively.
$K_{A}$ expresses the value of the antisymmetric part of the diffusion tensor, where its explicit form results from the effects on the motion
of \GCR particles due to drift.
$\vec{V}_{\rm{sw}}$ is the SW speed and $\vec{V}_{D}$ is the guiding center speed for a pitch angle-averaged nearly isotropic distribution function.
The equation can be then re-written as:
\begin{equation}
  \frac{\partial f}{\partial t}  -  \nabla\cdot [\mathbf{K}^{S}\cdot\nabla f ]
    +  (\vec{V}_{\rm{sw}} + \vec{V}_{D}) \cdot\nabla f 
   - \frac{(\nabla \cdot\vec{V}_{\rm{sw}})}{3}\frac{\partial f}{\partial (\ln\!{R})}  = 0 \,,
\end{equation}
The motion of the \GCR particles in the HMF is usually decomposed in a regular gradient-curvature and HCS drift motion on the background
average HMF and a diffusion due to the random motion on the small-scale fluctuations of the turbulent HMF.  
All these effects are included in the diffusion tensor $\mathbf{K}$ of Eq.(\ref{diff-tens}), which can be decomposed
in a symmetric part that describes the diffusion and an antisymmetric one that describes the
drifts, \ie, $\mathbf{K}=\mathbf{K}^S+\mathbf{K}^A$, with $K_{ij}^S = K_{ji}^S$ and $K_{ij}^A = -K_{ji}^A$.
Particle moving in a magnetic turbulence are pitch-angle scattered by the random HMF irregularities. 
This process is captured by the symmetric part of the diffusion tensor $\mathbf{K}^S$, which is diagonal if the $z$-coordinate is aligned with the background HMF.
Three diffusion coefficients are therefore needed, namely, parallel diffusion $K_{\parallel}$, transverse radial, $K_{\perp{r}}$, and transverse
polar diffusion coefficient $K_{\perp\theta}$. The coefficients can also be expressed in terms of mean free path
$\lambda$ along the background HMF, \eg, $K_{\parallel} = \beta c \lambda_{\parallel}/3$ (with $\beta= v/c$).
The determination of the diffusion coefficients is a key ingredient to study the propagation of charged particles
in turbulent magnetic fields like the HMF and is the subject of many theoretical and computational studies.
The Quasi Linear Theory (QLT) has been successful at describing parallel diffusion, especially in its time-dependent and non-linear extensions \citep{Jokipii1966QLT}. 
Regarding perpendicular diffusion, the QLT provides upper limits within the field line random walk
description \citep{Jokipii1966QLT,Giacalone1999QLT}, while the best approaches
follow the nonlinear guiding center theory \citep{Matthaeus2003,Shalchi2004,Shalchi2020}.

From a microscopic point of view, \GCR diffusion is linked to the resonant scattering of particles with rigidity $R$
with the HMF irregularities around the wave number $k_{\rm{res}} \sim 2\pi /r_{L}$, where $r_L =R/B$.
The essential dependence of $\lambda_{\parallel}$ on the HMF power spectrum can be expressed
as $\lambda_{\parallel} \sim r^2_L \langle B^2\rangle/w(k_{\rm{res}}) \sim R^2/w(k_{\rm{res}})$,
where $\langle B^2\rangle$ is mean square value of the background field and $w(k_{\rm{res}})$
is the power spectrum of the random fluctuations of the HMF around the resonant wave number.
The power spectral density follows a power-law as $w(k) \sim k^{-\nu}$, where the index $\nu$ depends on
the type and on the spatial scales of the turbulence energy cascade \citep{Kiyani2015,Bruno2017}.
Therefore, $\lambda_{\parallel}$ depends on the turbulence spectral index as $\lambda_{\parallel} \sim R^{2-\nu}$
In this work, for the rigidity and spatial dependence of the parallel diffusion coefficient,
we adopt a double power-law rigidity dependence
and an inverse proportionality with the local HMF magnitude, following \citem{Potgieter2014}: 
\begin{equation}\label{Eq::KvsR}
K_{\parallel} = K_{0}\frac{{\beta}}{3} \frac{(R/R_{0})^{a}}{(B/B_{0})} \left[ \frac{(R/R_0)^h + (R_k/R_0)^h }{1 + (R_k/R_0)^h} \right]^{\frac{b-a}{h}} \,.
\end{equation}
In this expression, $K_{0}$ is a constant of the order of $10^{23}$ cm$^2$s$^{-1}$,  $R_{0}$\,=\,1\,GV to set the rigidity units, $B$ the HMF magnitude and $B_{0}$
the field value at Earth and written in a way such that the units are in $K_0$. Here $a$ and $b$ are power indices  that determine the slope of the
rigidity dependence, respectively,  below and above a rigidity $R_k$, whereas $h$ determines the smoothness of the transition.
The perpendicular diffusion in the radial direction is calculated as $K_{\perp r} = \xi_{\perp r} \times K_{||} $,
while the polar perpendicular diffusion was parameterized as $K_{\perp\theta} =  \xi_{\perp \theta}\times g(\theta)\times K_{||} $,
where $g(\theta)$ is a function that enhances $K_{\perp\theta}$ by a factor $d$ near the poles, defined as \citep{Potgieter2014}:
\begin{equation}
g(\theta) =  A^+ \mp A^-\tanh \left[   8\left(  \theta_A + \pi/2 \pm \theta_F  \right)   \right] \,. 
\label{polar_enhancement}
\end{equation}
Here $A^{\pm} = (d\pm 1)/2$, $\theta_F = 35^{\circ}$ and $\theta_A = \theta$ if $\theta\leq \pi/2$ or $\theta_A = \pi - \theta$ if $\theta\geq \pi/2$, with $d$ = 3. 
The enhancement in the latitude direction of $K_{\perp\theta}$, together with the anisotropy between the perpendicular diffusion
coefficients and HMF modification at the polar regions,  is needed to account for the very small latitudinal dependence of
the \GCR intensity, as it was observed in the Ulysses data \citep{Potgieter1989,Ulysses1}.
The adoption of constant $\xi_{\perp}$-factors implies that $K_{\perp}$ and $K_{\parallel}$ follow the same rigidity dependence, which
may be a simplification in the high-$R$ domain \citep{Shalchi2004,Qin2002QLT}.
Nonetheless, QLT-based simulations agree for nearly rigidity-independent $\xi$, with the typical value of 0.02-0.04 \citep{Giacalone1999QLT,Hussein2015}.
In this work, the parameters $\xi_{\perp r}$ and $\xi_{\perp\theta}$ are fixed to the value 0.02.
We now turn on drift effects, that account for the charge-sign and polarity dependence of \GCR transport in the HMF \citep{JokipiiThomas1981,Webber2005}.
The regular motion of \GCRs on the large-scale HMF is given by the pitch-angle averaged guiding center drift speed $\langle\vec{V}_{D}\rangle$.
It can be related to the antisymmetric part of the diffusion tensor \citep{Burger1995}:
\begin{equation}
\langle(V_{D})_i\rangle = \frac{\partial K_{ij}^A}{\partial x_j} \,,
\end{equation}
where the antisymmetric part of the tensor has the form:
\begin{equation}
\label{antisym0}
K^{A}_{ij} = K_A u(\theta)\zeta(R)\epsilon_{ijk}\frac{B_k}{B} \,.
\end{equation}
Here $\epsilon_{ijk}$ is the Levi-Civita symbol, $u(\theta)$ is a function that describes the transition between the region influenced by the HCS
and the regions outside of it and $\zeta(R)$ is a function of rigidity that suppresses drifts at low rigidity.
To determine the value of $K_A$, we note that the small value of the ratio $K_{\perp}/K_{\parallel}$ suggests that \GCR particles
move over many gyro-orbits in a mean free path, therefore the drift motion is weakly affected by scattering. 
In the weak scattering approximation, one has:
\begin{equation}
\label{antisym}
K_A = K_{A}^{0}  \frac{Q}{|Q|}\frac{\beta R}{3B} \,,
\end{equation}
where $Q$ is the \GCR particle charge and $K_{A}^{0}$ is a normalization factor $\leq$ 1.
Drift motion is relevant close the HCS, where \GCRs cross many times regions of opposite HMF polarity.
A 2D description of HCS drift is given in \citet{Burger1995}.
In this approach, the drift velocity is given by: 
\begin{equation}
\label{drift}
  \langle \vec{V}_{D} \rangle = \zeta(R)  \left[ \vec{G} + \vec{H} \right] \,,
\end{equation}
where the two vectors are defined as follows:
\begin{equation}
\label{drift2}
\begin{aligned}
  \vec{G}  &= u(\theta) \nabla \times \left( K_A \frac{ \vec{B} }{ B } \right) \\
  \vec{H}  &= \left( \frac{ \partial u(\theta) } {\partial \theta} \right) \left( \frac{ K_{A} }{ r } \right) \vec{e_{\theta}} \times \frac{ \vec{B} }{ B } 
\end{aligned}
\end{equation} 
The $\vec{G}$-term in Eq.(\ref{drift2}) describes the gradient-curvature drifts, the $\vec{H}$-term describes the
particle motion across the region affected by the HCS, $\vec{e}_{\theta}$ is the unit vector along the polar direction, and $u(\theta$) is given by:
\begin{equation}
 u(\theta) =
  \begin{cases}
     (1/a_h)\arctan\{  \left[   1 - \left( 2\theta/\pi\right)\tan a_h  \right]    \} &  \text{if } c_h < \pi/2\\
     1 - 2H(\theta - \pi/2)  & \text{if } c_h = \pi/2
  \end{cases}
\end{equation}
with $H$ the Heaviside step function,
\begin{equation}
 a_h = \arccos \left( \frac{\pi}{2c_h} - 1 \right) \,,
\end{equation}
and
\begin{equation}
 c_h = \frac{\pi}{2} - \frac{1}{2}\sin \left( \alpha + \frac{2r_L}{r} \right) \,.
\end{equation}
The angle $2r_L/r$ depends on the maximum distance that a particle can be away from the HCS while drifting.
Finally, the function $u(\theta)$ is such that $u(\pi/2)$ = 0, $u(c_h)$ = 0.5 and $\partial u (\pi/2)/\partial\theta$ = 1.
\GCR drift coefficients are expected to be reduced in presence of turbulence as results theoretically and from numerical test-particle simulations \citep{Tautz2015,Engelbrecht2017}.
In this work, we use a simple approach to incorporate drift reduction.
Following \citem{Engelbrecht2017}, we adopt a reduction factor of the type:
\begin{equation}
\label{drift_red1}
\zeta = \frac{1}{1 + \frac{R_{A}^{2}}{R^{2}}} \,,
\end{equation}
where the reduction occurs at rigidity below the cutoff value $R_{A}=\lambda_{\perp}\delta B_{T}$, which depends on
the perpendicular diffusion length and total variance of the HMF. 
The reduction is effective at $R \ll R_{A}$, when $\zeta \approx (R/R_A)^2 \ll 1$, while in the high-$R$ limit one has $\zeta \approx 1$.
The cut-off value $R_{A}$ depends on the HMF turbulence through $\lambda_{\perp}$ and $\delta B_{T}$.
With typical values of $\lambda_{\perp} \approx 1.5\times 10^{-3}$ AU and $\delta B_{T} \approx 3.5$\,nT for the considered epochs,
one can estimate $R_{A}\approx$\,0.3$\,\-\,$0.6 GV. In this work we have fixed it at 0.5 GV, corresponding to a proton kinetic energy of 125 MeV.
The normalization $K_A^0$ factor is fixed to 1, so that the whole drift reduction is regulated by $\zeta$.

The most relevant feature of magnetic drift is that its direction depends on the $\it{sign}$ of the charge, $\hat{q}=Q/|Q|$,
and on the HMF polarity $A$, via the product $\hat{q}A$, so that particles with opposite $\hat{q}A$ will drift in
opposite directions and will follow different trajectories in the heliosphere.
This characteristic is expected to give observable charge-sign dependence in the \GCR modulation.
Finally, in a reference frame with the z coordinate along the average magnetic field, the diffusion tensor is given by Eq.(\ref{diff-tens}).
The effective diffusion tensor in heliocentric polar coordinates is obtained by a coordinate transformation in the modified Parker's field. 
In our 2D approach, the relevant components are
$K_{rr} = K_{\parallel}\cos^{2}\psi + K_{\perp{r}}\sin^2\psi$, $K_{\theta\theta} =  K_{\perp\theta}$ and $K_{\theta{r}} = K_{A}\sin\psi = -K_{r\theta}$.
%
\begin{figure}[t]
\centering
\includegraphics[clip,trim=.05cm 0.05cm 0.05cm 0.05cm, width=0.4\textwidth,scale=0.30]{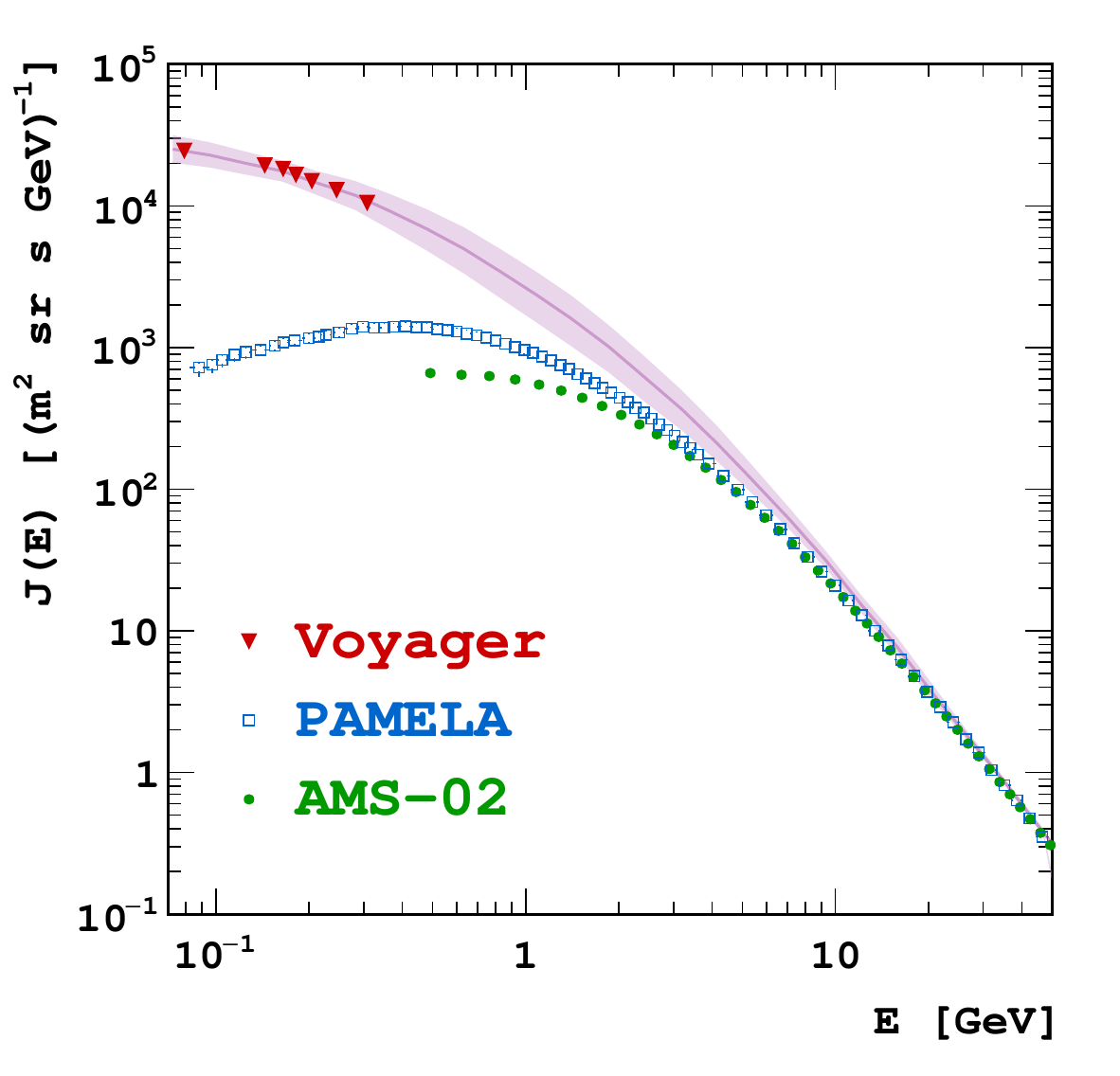}
\caption{Proton LIS used as input boundary condition for the modulation along with the estimated uncertainty band \citep{Tomassetti2015TwoHalo,Feng2016,Tomassetti2018PHeVSTime,Tomassetti2019Numerical}. 
  Data from Voyager-1 in interstellar space, and from AMS-02 and PAMELA in low Earth orbit collected during two epochs.} %
\label{Fig::ccProtonLIS}
\end{figure}

\subsection{The proton LIS} 
\label{Sec::ProtonLIS}      
%
To resolve the modulation equation for cosmic protons, their LIS must be specified as boundary condition.
The determination of the \GCR proton LIS requires a dedicated modeling effort, starting from the distribution of Galactic \GCR sources
and accounting for all the relevant physical processes that occur in the interstellar medium.
In this work, we adopt an input LIS for \GCR protons that relies on a \emph{two-halo model} of \GCR{}
propagation in the Galaxy \citep{Tomassetti2015TwoHalo,Feng2016}. 
In this model, the injection of primary \GCRs in the ISM is described by rigidity-dependent source terms
$S\propto(R/{\rm GV})^{-\gamma}$ with $\gamma=$\,2.28$\pm$0.12 for protons.
The diffusive transport in the $L$-sized Galactic halo is described by an effective diffusion coefficient
$D = \beta D_{0}(R/GV)^{\delta_{i/o}}$  with  $D_{0}/L=0.01\pm$0.002\,kpc/Myr \citep{Feng2016,Tomassetti2017BCUnc}. 
The two spectral indices ${\delta_{i/o}}$ describe two different diffusion regimes in the inner/outer halo,
with $\delta_{i}=0.18\pm$0.05 for $|z|<\xi\,L$ (inner halo), and  $\delta_{o}=\delta_{i}+\Delta$ for $|z|>\xi\,L$ (outer halo), with $\Delta=0.55\pm$0.11.
The $z$ variable here is the vertical spatial coordinate. The half-thickness of the halo is $L\cong{5}$\,kpc and the near-disk region (inner halo)
is set by $\xi=0.12\pm$0.03. Finally, we considered the impact of diffusive reacceleration. 
Within the two-halo model, the interstellar Alfv{\'e}nic speed is constrained from the data to lie between 0 and 6\,km\,s$^{-1}$. 
Calculations of the proton LIS were constrained by various sets of measurements: low-energy proton data (at 140\,--\,320 MeV) collected by Voyager-1
beyond the HP, high-energy proton measurements ($E\gtrsim$\,60 GeV) made by AMS-02 in low Earth orbit,
along with measurements of the B/C ratio from both experiments. The latter were essential to constrain the
diffusion parameters of the LIS model \citep{Tomassetti2017BCUnc}.
Details on this model are provided elsewhere \citep{Feng2016,Tomassetti2018PHeVSTime}.  
The resulting proton LIS is shown in Fig.\,\ref{Fig::ccProtonLIS} in comparison with the data from Voyager-1, along with
PAMELA and AMS-02 measurements made in March 2009 and April 2014, respectively.
The uncertainty band associated with the calculations is also shown in the figure.
This model is in good agreement with other recently proposed LIS models \citep{Boschini2018, Corti2019, Corti2016, Tomassetti2017TimeLag, Tomassetti2015PHeAnomaly}.

\section{Data Analysis}     
\label{Sec::Fitting}        

In this section, we present the analysis method by which we extract knowledge and insights from the data
using the mathematical framework described Sect.\,\ref{Sec::Model}. 
In practice, we defined a set of physics observables, to be computed as model predictions,
and a set of model parameters to be determined by statistical inference.

\subsection{The cosmic ray data} 
\label{Sec::TheData}             

The data used in this work consist in
time-resolved and energy-resolved measurements of \GCR proton fluxes, in the kinetic energy range from $\sim$\,80 MeV to $\sim$\,60 GeV.
Specifically, we use the 79 BR-averaged fluxes measured by the AMS-02 experiment in the International Space Station from May 2011 to May 2017 \citep{Aguilar2018PHeVSTime},
and the 47+36 BR-averaged fluxes observed by the PAMELA instrument in the satellite Resurs-DK1 from June 2006 to January 2014\,\citep{Martucci2018,Adriani2013Protons}.
%
\begin{figure*}[ht!]
\centering
\includegraphics[width=0.9\textwidth,scale=0.45]{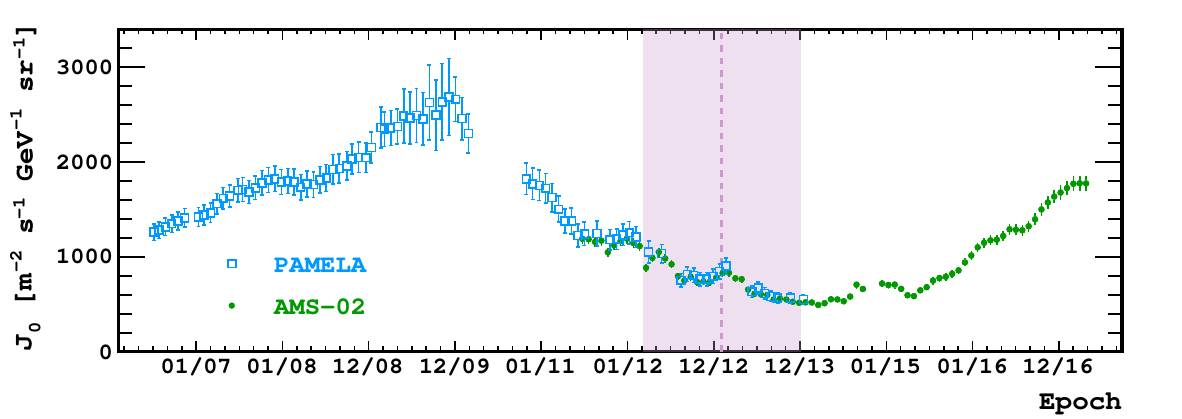}
\caption{BR averaged flux $J_{0}$ evaluated in the reference energy range between 0.49-0.62 GeV
  from PAMELA (open squares)\citep{Adriani2013Protons,Martucci2018}
  and AMS-02 (filled circles)  \citep{Aguilar2018PHeVSTime, Aguilar2018LeptonsVSTime}.
  The vertical dashed line shows the epoch of the HMF polarity inversion, along with the shaded area indicating the reversal epoch.}
\label{Fig::ReferenceFlux}       
\end{figure*}
%
The data sample corresponds to a total of 10,101 data points 
collected over a time range of  about 11\,years, from the solar minimum from 2006 to 2009, the ascending phase to solar maximum, 
when the HMF polarity $A$ reversed from $A<$0 to $A>$0, and the following descending phase until May 2017.
These data have been retrieved by the ASI-SSDC \emph{Cosmic Ray Data Base} \citep{DiFelice2017}. 

The intensity of the \GCR proton fluxes in the energy range between 0.49 - 0.62 GeV
are shown in Fig.~\ref{Fig::ReferenceFlux} as a function of time for both the PAMELA and AMS-02 data sets. 
From the figure, the complementarity of the two experiments is apparent.
It can be seen that the highest intensity of the \GCR is reached during $\sim$ December 2009, \ie, under the solar minimum,
while the lowest intensity occurs in $\sim$ February 2014, around solar maximum.
The vertical dashed line of the figure shows the HMF reversal epoch $T_{\rm{rev}}$, along with
the transition region shown as a shaded area where the HMF is disorganized and the polarity is not defined.
The determination of $T_{\rm{rev}}$ and the transition region are presented later on. 

\subsection{The parameters} 
\label{Sec::TheParameters}  

The numerical model presented in Sect.\,\ref{Sec::Model} makes use of several physics input to be determined with the help of observations.
Inputs include \emph{solar parameters}, characterizing the conditions of the Sun or the interplanetary plasma,
and \emph{transport parameters} that describe the physical mechanisms of \GCR propagation through the plasma.
Solar and transport  parameters are inter-connected each other and they may show temporal variations related to the solar cycle.
For instance, solar parameters such the magnetic field magnitude, its variance and its polarity 
are transported from the Sun into the outer heliosphere, therefore provoking time-dependence \GCR diffusion and drift.

We identified, in our model, a set of six time-dependent key parameters that are of relevance for the phenomenology of \GCR modulation.
They are the tilt angle of the HCS $\alpha(t)$, the strength of the HMF at the Earth's location $B_{0}(t)$, the HMF polarity $A(t)$,
and the three diffusion parameters appearing in Eq.(\ref{Eq::KvsR}): the normalization factor
of the parallel diffusion tensor, $K_{0}(t)$, and the two spectral indices of the rigidity-dependence
of \GCR diffusion, $a(t)$ and $b(t)$, below and above the break $R_{k}$, as seen in Eq.(\ref{Eq::KvsR}).
Note that all key parameters are expressed as continuous functions of time $t$, but 
in practice, they have been determined for the epochs corresponding to the \GCR flux measurements.
%
\begin{figure*}[hbt!]
\centering
\includegraphics[width=1.0\textwidth,scale=0.50]{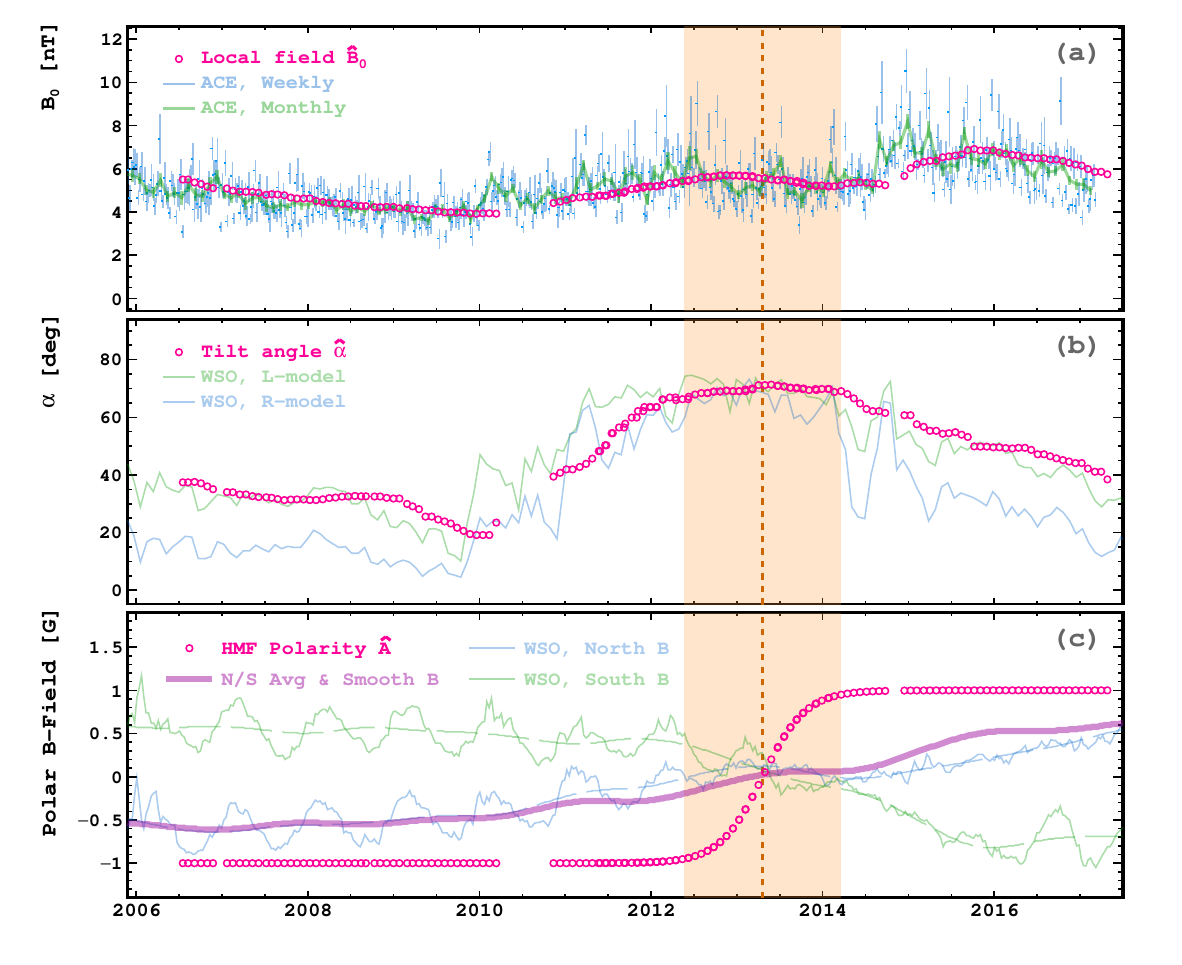} 
\caption{Reconstruction of the tilt angle $\alpha$, the local HMF strength $B_{0}$, and the magnetic polarity $A$ as a function of the epoch,
  evaluated with the BMA procedure in correspondence of the epochs of the PAMELA and AMS-02 flux measurements.
  The vertical dashed line marks the HMF reversal $T_{\rm{rev}}$. The shaded area around $T_{\rm{rev}}$ represent the effective period of the HMF polarity transition.
  The raw data, shown as thin shaded lines, are taken from the ACE space probe and from the Wilcox Solar Observatory \citep{Hoeksema1995,Smith1998}.} 
\label{Fig::SolarParameters}       
\end{figure*}

The three solar parameters ${\alpha, B_{0}, A}$ can be determined from solar observatories:
data of HMF polarity and tilt are provided by the Wilcox Solar Observatory on 10-day or BR basis.
Measurements of the HMF $B_{0}$ at 1 AU are done \emph{in-situ} on daily basis, since 1997,
by the Advanced Composition Explorer (ACE) on a Lissajous orbit around $L1$ \citep{Smith1998}.
It is important to notice that, in this study, our aim is to capture the effective status of the large-scale heliosphere
sampled by \GCRs detected at a given epoch $t$, and this is connected to solar-activity parameters that are \emph{precedent} to that epoch.
In fact, several studies have reported a time lag of a few months between the solar activity  and the
varying \GCR fluxes \citep{Tomassetti2017TimeLag,SierraPorta2018}, reflecting the fact that the perturbations
induced by the Sun's magnetic activity take a finite amount of time to establish their effect in the heliosphere.
To tackle this issue, for each epoch $t$ associated to a given \GCR flux measurement,
we perform a Backward Moving Average (BMA) for $\alpha$ and $B_{0}$, and $A$, \ie, a time-average of these quantities calculated
over a time window $[t-\tau, t]$.
The window extent $\tau$ is the time needed by the SW plasma to transport the magnetic perturbations from the Sun to the HP boundary,
which ranges between $\sim$\,8\,months (fast SW speed) and $\sim$\,16\,months (slow SW speed).
In the case of $\alpha$, the window is large because the HCS is always mostly confined in the slow (equatorial) SW region.
In the case of $B_{0}$, the BMA has to be computed by an integration over the latitudinal profile of the SW speed at a given epoch.
Our estimations are consistent with the lag reported in other studies \citep{Tomassetti2017TimeLag,SierraPorta2018} 
and supported by correlative analysis that we made \emph{a posteriori}.
Figure\,\ref{Fig::SolarParameters} shows the reference parameters $\widehat{B}_{0}$, $\widehat{\alpha}$ calculated for 
for each reference epoch $t$ corresponding to a BR-averaged \GCR measurement. A similar estimate is done for the polar magnetic field and for the
resulting polarity $\widehat{A}$, in Fig.\,\ref{Fig::SolarParameters}d.
The latter can be regarded as a ``smoothed'' definition for the magnetic polarity $A$, otherwise dichotomous ($A$=$\pm{1}$).
When the HMF is in a defined polarity state, one has $\widehat{A}=\pm{1}$.
During the HMF reversal transition epoch (shaded area in the figures), as the polarity is not well defined,
the estimate of $\widehat{A}$ takes a floating value between $-1$ and $+1$.
%
%
\begin{figure}[!]
\centering
\includegraphics[width=0.40\textwidth,scale=0.30]{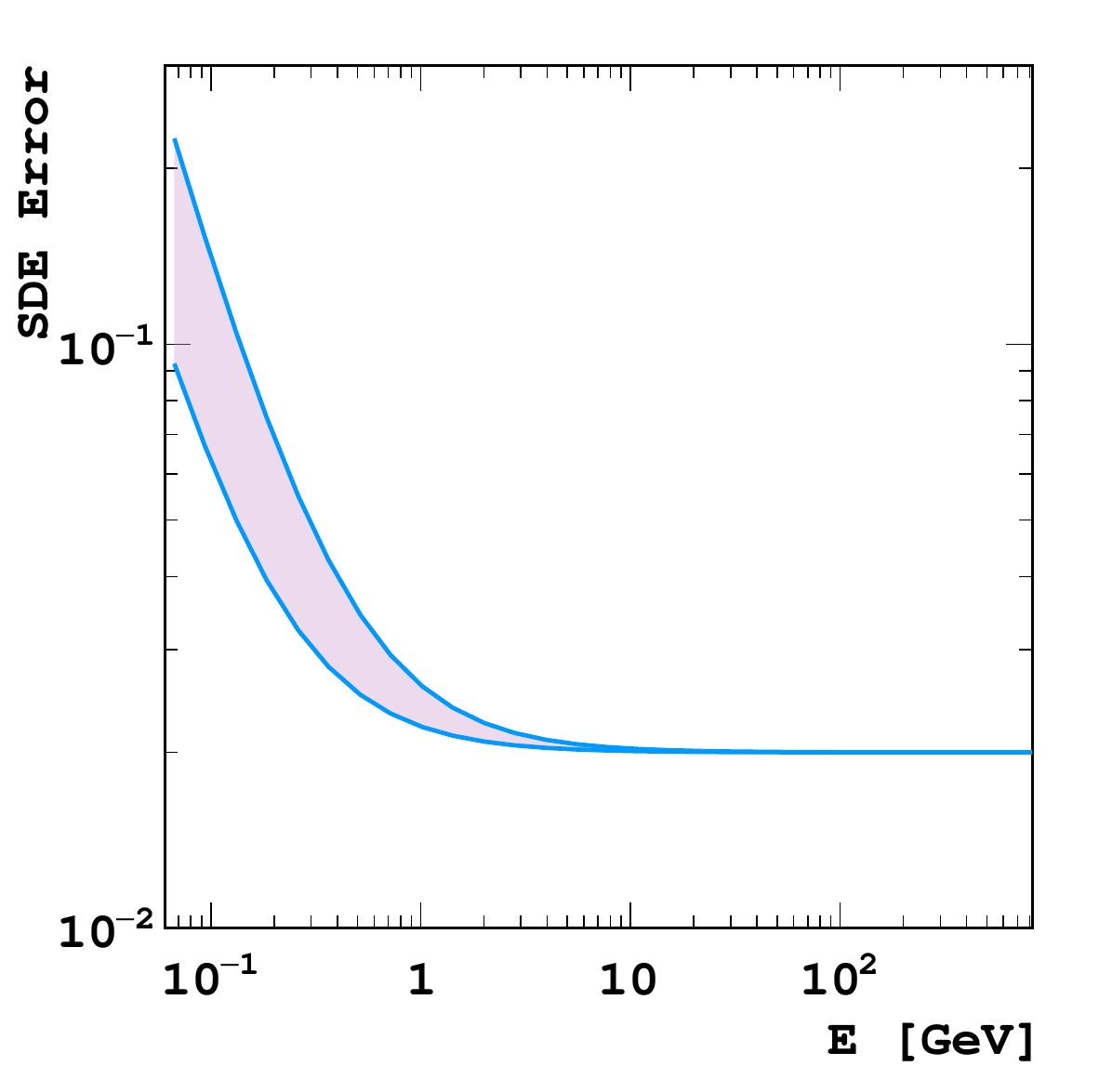}
\caption{Error band reflecting the statistical uncertainty of the Monte Carlo generated trajectories.
  Low-energy \GCRs have less chances to reach the inner heliosphere, giving a higher uncertainty on the Monte Carlo statistics.}
\label{errsde}       
\end{figure}
%
%
%
\begin{figure*}[hbt!]
\centering
\includegraphics[width=0.94\textwidth,scale=0.50]{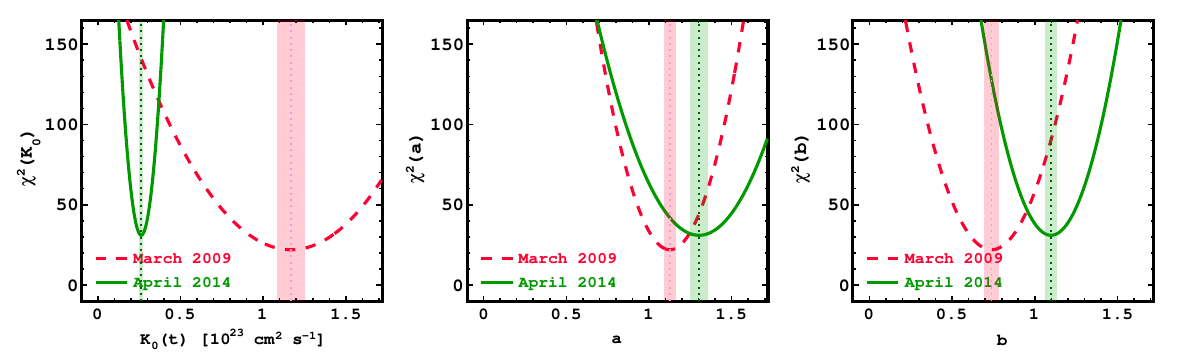} 
\caption{%
  One-dimensional projections of the $\chi^{2}$ surfaces as function of the transport parameters $K_{0}$, $a$, and $b$ evaluated
  for two epochs:  March 2009 (pink dashed line)  and April 2014 (green continuous line).
  In the two epochs, representing solar minimum and solar maximum conditions, the \GCR flux data come from PAMELA and AMS-02, respectively.}
\label{Fig::Chi2Surface}       
\end{figure*}

At this point, we also recall that several parameters entering the model that have been kept constant in the simulation, \ie,
assumed to be known or time-independent.
The HP and TS positions were fixed at $r_{\rm{HP}}$=122 AU and  $r_{\rm{TS}}$=85 AU, deduced from the Voyager-1 observations.
The data suggest that the TS may vary over the solar cycle of the order of a few AU, but its impact in the \GCR fluxes is
not negligible \citep{Tomassetti2017TimeLag}.

The $h$ parameter of Eq.(\ref{Eq::KvsR}), describing the smoothness of the transition between the
two diffusion regimes below and above $R_{k}$, was kept constant at $h=3$.
Within the precision of the data, the $h$ parameter has no appreciable impact on the \GCR fluxes.
Similarly, the rigidity break  $R_{k}$ for $K_{\parallel}$  was kept fixed at the value 3\,GV. 
This parameter represents the scale rigidity value where the \GCR Larmor radius matches the correlation length of
the HMF power spectrum, which is at the GV scale.
Regarding the value of $R_{k}$, we found that time variations on this quantity do not give appreciable variations
in the \GCR fluxes \citep[see,\eg][]{Potgieter2014}.
The $\xi_{\perp{i}}$ coefficients for the diffusion tensor,
for which the values used here represent a widely used assumption \citep[\eg,][]{Potgieter1989}. 
The polar enhancement factor of Eq.(\ref{polar_enhancement}) is kept constant at $d=3$
for $\xi_{\perp\theta}$ so that the condition $K_{\perp}/K_{\parallel} \ll 1$ is still fulfilled at the polar regions.
Regarding magnetic drift, the critical rigidity $R_{A}$ of Eq.(\ref{drift_red1}) is kept constant at 0.5 GV following previous
studies and independent observations on the \GCR latitudinal gradient \citep{Potgieter2014,Minnie2007}. 
This choice could be tested only with low-rigidity \GCR data ($R\ll{R_{A}}$), as our results are insensitive to the exact value of $R_{A}$. 
The normalization factor for drifts speeds $K_{A}^{0}$ was chosen to be unity such to set ``full drift'' speeds in the
propagation model for all the periods, and this the drift reduction is entirely given by Eq.(\ref{drift_red1}). 
Reductions in the $K_{A}^{0}$-value may occur during periods of strong magnetic turbulence, \eg, during solar maximum \citep{Minnie2007,Ferreira2004}.

\subsection{The statistical inference} 
\label{Sec::TheInference}              

\MyTitle{The parameter grid} 
%
The transport parameters $K_{0}(t)$,  $a(t)$ and $b(t)$ have been determined from the AMS-02 and PAMELA data by means of a global fitting procedure.
For this purpose a six dimensional discrete grid of the model parameters vector $\vec{q}=$ ($\alpha$, $B_0$, $A$, $K_0$,  $a$, $b$)  was built,
\ie, the model was run for every node of the grid such to produce a theoretical calculation for the \GCR proton flux.
In the grid, the parameter $\alpha$ ranges from $5^{\circ}$ to $75^{\circ}$ with steps of $10^{\circ}$,
$B_{0}$ from 3 to 8\,nT with steps of 1\,nT, and the polarity $A$ takes the two values $A=+1$ and $A=-1$.
The parameter $K_{0}$ ranges from 0.16 to 1.5 $\times$\,$10^{23}$ cm$^2$s$^{-1}$, with steps of 0.08 $\times$\,$10^{23}$ cm$^2$s$^{-1}$,
the indices $a$ and $b$ range from 0.45 to 1.65 with steps of 0.05. The total number of grid nodes amounts to 938,400.
For each node of the parameter grid, a theoretical prediction for the modulated proton flux $J_{m}(E, \vec{q})$ was evaluated,
as function of kinetic energy, over 120 energy bins ranging from 20\,MeV to 200\,GeV with log-uniform step. 
Using the SDE technique, $2\times 10^{3}$ pseudo-particles were Monte Carlo generated and retro-propagated for each energy bin.
This task required the simulation of about 14 billion trajectories of pseudo-protons, corresponding to several months of CPU time. 
Once the full grid was completed, the output flux was tabulated and properly interfaced with the data.
For each data set $J_{d}(E,t)$, representing a set flux measurements as function of energy for a given epoch $t$,
a $\chi^{2}$ estimator was evaluated as:
\begin{equation}
\label{chi2}
\chi^{2}(\vec{q} )= \sum_{i}  \frac{\left[ J_{d}(E_{i},t) - J_{m}(E_{i}, \vec{q}) \right]^{2} }{\sigma^2(E_{i},t)}  \,.
\end{equation} 
Similarly to $J_{m}$, the $\chi^{2}$ estimator is built such to be a continuous function of the parameters $\vec{q}$,
except for the variable $A$ that is treated as discrete.
From the $\chi^{2}$ estimator, the transport parameters $\{K_{0},a,b\}$ can be determined by minimization at any epoch, 
while the solar parameters $\{B_{0},\alpha,A\}$ can be considered as ``fixed inputs'', as they are
determined by the epoch $t$ using the BMA reconstruction presented above. 
For a given set of BMA inputs such as $\widehat{B}_{0}$ and $\widehat{\alpha}$, the flux $J_{m}(E,\vec{q})$ can be expressed as
a continuous function of the parameters by means of a multilinear interpolation over the grid nodes.
In the $\alpha-B_{0}$ plane, one has $\alpha_{j}<\widehat{\alpha}(t)<\alpha_{j+1}$ and $B_{0k}<\widehat{B}_{0}(t)<B_{0,k+1}$,
where $\alpha_{j}$ and $B_{0,k}$ are the closest values of the grid corresponding to their BMA averages.
Regarding polarity $A$, both $\pm{1}$ evaluations were done under the assumption that the polarity is known.
The flux model dependence upon energy should also be handled. In Eq.(\ref{chi2}), $E_{i}$ are the mean measured energies
reported from the experiments (coming from binned histograms). In general, the $E_{i}$ array does not correspond to the energy
grid of the model. The model evaluation of $J_{m}(E,\vec{q})$ at the energy $E_{i}$ was done by log-linear interpolation.

\MyTitle{The uncertainties}  
%
The $\sigma$ factors appearing in Eq.(\ref{chi2}) represent the total uncertainties associated with the flux.
They can be written as $\sigma^{2}(E_{i},t) = \sigma_{d}^{2}(E_{i},t) + \sigma_{m}^2(E_{i},t)$.
Here $\sigma_{d}^{2}(E_{i},t)$ are the experimental errors associated to the flux measurement of the $i$-th energy bin around $E_{i}$,
while $\sigma_{\rm{m}}^2(E_i,t)$ are the theoretical uncertainties of the flux calculations evaluated at the same value of energy.
Uncertainties in experimental data are of the order of $10\,\%$ in the PAMELA data and $\sim\,2\%$ in the AMS-02 data, 
although they depend on kinetic energy.
Theoretical uncertainties include statistical fluctuations of the finite SDE generation of pseudo-particle trajectories.
Uncertainties are relevant at low energy where, due to the heavy adiabatic energy losses,
the Monte Carlo sampling suffers from a smaller statistics.
Thus, after repeating many times the simulation with the same modulation parameters, the modulated flux will fluctuate around
an average value because of the random process of pseudo-particles propagation with the SDE approach.
These fluctuations can be arbitrarily reduced with the increase of the pseudo-particle generation, but at the expense of a large CPU time.
The evaluation of these uncertainties can be done as follows.
Given $N_{m}$ as the number of pseudo-particles that reach the boundary with energy $E$, and $N_{G}$ as the number of pseudo-particles
generated at the same energy, the ratio of the modulated flux to the LIS flux is $J_{m}/J_{\rm{LIS}} \approx N_{m}/N_{G}$.
Since the propagation process is stochastic in nature, the relative error of the modulated flux scales as $\delta J_{m}/J_{m} = 1/\sqrt{N_{m}}$, where $N_{m} = N_G(J_{m}/J_{\rm{LIS}})$.
We found that the generation of $N\cong{2\times10^{3}}$ pseudo-particles for each energy bin is sufficient for being not dominated by SDE-related uncertainties.
The relative uncertainties as function of kinetic energy are shown in Fig.~\ref{errsde}.
The errors are about $\sim\,10\,-\,20\%$ at 20 MeV of energy and decrease with increasing energy. They become constant at $\sim\,2\%$ above few GeVs.
A minor source of systematic error comes from the multilinear interpolation of the parameter and energy grid,
\ie, from the method we used to evaluate the flux at any arbitrary set of parameters and energy.
From dedicated runs, we have estimated that the uncertainty introduced by the interpolation,
rather than the direct simulation with of $J(\vec{q}, E)$, is always of the order of 1\,\%.
An important source of systematic error is the uncertainty coming from the input LIS of \GCR protons, see Sect.\,\ref{Sec::ProtonLIS}.
The LIS uncertainties are highly energy-dependent. They are significant in the energy region of $\sim$\,1-10 GeV (up to 30\,\% and more), 
where direct interstellar data are not available but the modulation effect is still considerable. 
However, in this energy region, the Galactic transport parameters regulating the LIS intensity are in degeneracy
with the free parameters of \GCR diffusion (Sect.\,\ref{Sec::TheParameters}) and in particular with $K_{0}$ \citep{Tomassetti2018PHeVSTime}.
Such a degeneracy translates into a correlation between the best-fit $K_{0}$ values and the LIS intensity at the GeV scale which, in turn, 
determines the absolute scale of the the modulated \GCR flux $J_{0}$ at the GeV scale.
The $K_{0}-J_{0}$ correlation is also discussed in Sect.\,\ref{Sec::TemporalDependencies}.
To estimate the impact of the LIS uncertainty on the temporal dependence of the best-fit parameters of \GCR diffusion in heliosphere,
we proceeded as in \citem{Tomassetti2018PHeVSTime,Tomassetti2019Numerical}.
We performed dedicated runs of fitting procedure for a large number of randomly generated LIS functions where, for each input LIS,
the time-series of the diffusion parameters were determined. In practice, the LIS functions were generated using the Monte Carlo
framework in \citem{Feng2016}, \ie, according to the probability density function of the Galactic \GCR transport parameters.
With this procedure, the systematic uncertainties associated with the LIS modeling are included in the final errors with a proper account for their correlations.

\MyTitle{The reversal phase} 
%
The parameter $T_{\rm{rev}}$ marks the epoch of the 2013 magnetic reversal, where the HMF flipped from negative to positive polarity states
The polarity of the HMF, however, is well defined only for $t\ll$$T_{\rm{rev}}$ and $t\gg$$T_{\rm{rev}}$, where the large-scale HMF structure
follows a dipole-like Parker's field to a good approximation.
During reversal, the polarity of the field is less sharply defined and the HMF field follows a more complex dynamic \citep[\eg,][]{Sun2015}. 
A way to account for this situation is to use a generalized definition of polarity, such as the BMA
reconstruction $\widehat{A}$ of Fig.\,\ref{Fig::SolarParameters} which ranges from -1 to +1.
For any  given parameter configuration $q$, the flux model $J_{m}(E,\vec{q})$ can be built as a linear combination
of fluxes with defined polarities, weighted by a transition function $\mathcal{P}{\equiv}(1-\widehat{A})/2$:
\begin{equation}
J_{m}(E, \vec{q}) = J^{-}_{m}(E; \vec{q}^{-})\mathcal{P} + J^{+}_{m}(E; \vec{q}^{+})\left[1 - \mathcal{P}\right]  \,,
\label{weightedflux}
\end{equation}
where $\vec{q}^{(\pm)}$ = $\{\alpha,B_{0}, A^{\pm}, K_{0}, a, b\}$ is a vector of parameters with fixed polarity $A=\pm$1,
and $J^{(\pm)}_{m}$ are the corresponding modulated fluxes.
The weight $\mathcal{P}$ ranges from 1 to 0, for floating polarity $\widehat{A}$ ranging from -1 to 1. 
The time-dependence of the $\mathcal{P}(t)$-function associated to the polarity $\widehat{A}(t)$ of Fig.\,\ref{Fig::SolarParameters}
can be expressed as follows:
\begin{equation}
\mathcal{P}(t) = \left[ 1 + e^{\left( t-T_{\rm{rev}}\right)/\delta T} \right]^{-1}  \,,
\end{equation}
where $\delta T\,\cong$\,3\,months. 
The transition function $\mathcal{P}(t)$ is such that $\mathcal{P}\cong$\,0 ($\mathcal{P}\cong$\,1) for $t\lesssim\,3\,T_{\rm{rev}}$ ($t\gtrsim\,3\,T_{\rm{rev}}$)
within 1\,\% level of precision, \ie, when $t = T_{\rm{rev}}\pm 3\delta T$, the flux is 99\% made of a fixed polarity,
while the maximum mixing is for $t=T_{\rm{rev}}$ when  $\mathcal{P}(t)=$\,1/2.
It is worth noticing that Eq.(\ref{weightedflux}) relies on the implicit assumption that,
during HMF reversal, the modulated flux of \GCRs can be regarded as a superposition of fluxes with positive and negative polarity states.
We also note that this approach enabled us to define the transition epoch, from a smoothed definition of the polarity $\widehat{A}$,
which is indicated by the shaded area in Fig.\,\ref{Fig::SolarParameters}.
Such a definition of the transition epoch is consistent with estimations of the reversal epoch based on the dynamics of the HMF topology \citep{Sun2015,Pishkalo2019}.

\MyTitle{The parameter extraction} 
%
Our determination of the diffusion parameters $K_0(t)$, $a(t)$ and $b(t)$ is based on the least squares method.
In practice, we proceeded as follows.
Given a set of \GCR proton flux measurements $J_d(E,t)$, for each parameter $x$= $K_0(t)$, $a(t)$, and $b(t)$,
the corresponding $\chi^2(x)$ distribution, defined as in Eq.(\ref{chi2}), is evaluated.
The evaluation is done for all values of the other parameters $y\ne x$, marginalized over the hidden dimensions.
This returns a curve $\chi^{2}_{\rm{min}}(x)$ as function of the parameter $x$ and minimized over all hidden dimensions.
From the minimization of $\chi^{2}_{\rm{min}}(x)$, the best-fit  parameter $\hat{x}$ and its corresponding uncertainty are estimated.
For the minimization, we tested two approaches.
One method consisted in the interpolation with a cubic spline of the whole $\chi^2_{\rm{min}}(x)$ curve. 
A second method, similar to \citet{Corti2019}, consisted in the determination of the minimum $x_{i,\rm{min}}$ point from a parameter scan over the grid, and then by
making a parabolic re-fitting of the $\chi^{2}_{\rm{min}}(x)$ curve around the $x_{i,\rm{min}}$ and its adjacent points.
The position of the minimum and its uncertainty can be calculated as estimation of $x_{\rm{best}}$.
The errors on the parameters are estimated as $\sigma_{x}= max(|x_{-}  - x_{\rm{best}}|, |x_{+}  - x_{\rm{best}}|)$,
where $x_{\pm}$ is the parameter value such that $\chi^2_{\rm{min}}(x_{\pm}) = \chi^2_{\rm{min}}(x_{\rm{best}}) + 1$ above and  below $x_{\rm{best}}$,
which is the standard error estimation of the least squares method.
The little discrepancy of the two methods was used as a systematic errors which, however,
turned out to be negligible in comparison with the standard errors of the fit.
The shapes of the $\chi^{2}_{\rm{min}}$ projections as function of the diffusion parameters is illustrated in Fig.\,\ref{Fig::Chi2Surface}
for two distinct epoch March 2009 (BR\,2379 during solar minimum) and April 2014 (BR\,2466, during solar maximum).
For each curve, the best-fit parameter $\hat{x}$ is shown (vertical line) along with its associated uncertainty $\sigma_{x}$ (shaded band).
In the two considered epochs, the data come from PAMELA and AMS-02 experiment, respectively. As seen from the figure,
AMS-02 gives in general large $\chi^{2}$-values in comparison with PAMELA.
In both time series the convergence of the fit is good and the parameters are well constrained. It can be seen that the AMS-02 data provide
tight constraints on the $K_{0}$ and $b$ parameters, while the parameter $a$ is more sensitive to low-rigidity data and thus it is better constrained by PAMELA.
%
%
\begin{figure}[htb]
\centering
\includegraphics[width=0.40\textwidth,scale=0.30]{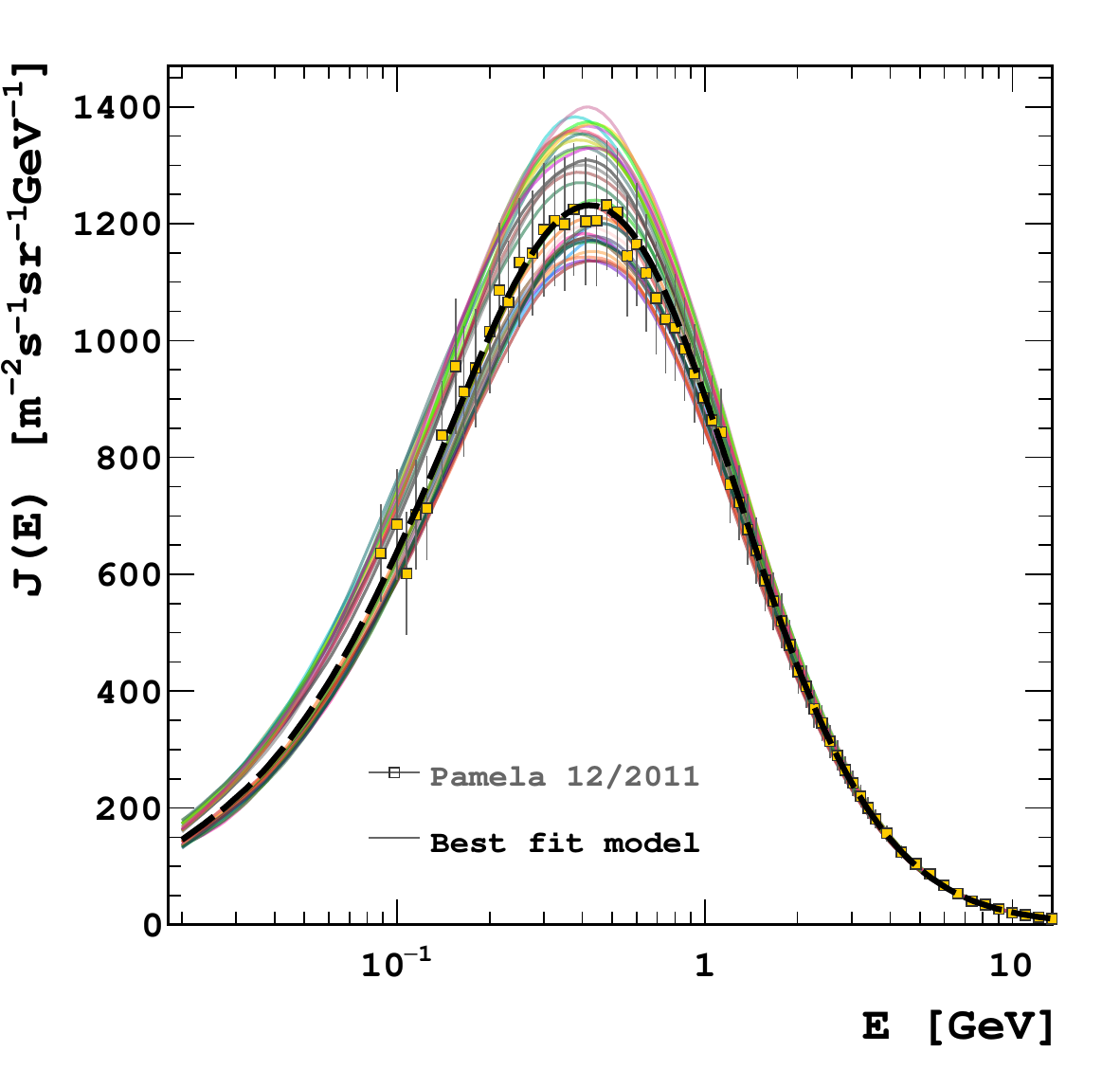}
\caption{Example of multilinear flux interpolation. The thick long-dashed line represents the interpolated best flux fit to the data set for 
  BR\,2435  - December 2011. The colored curves show the 32 fluxes corresponding to the array models such that $x_{i} \leq x_{best} < x_{i+1}$
  (where $x = \alpha, B_0, K_0, a$ and $b$t ) used for the interpolation.}
\label{interpol}       
\end{figure}
%
After the best-fit parameters have been determined for a give set of data, the best model flux $J_{\rm{best}}(E)$ is recalculated
using a multilinear interpolation over the 5-dimensional grid such that $x_i \leq x_{\rm{best}} < x_{i+1}$ where $x = \alpha, B_{0}, K_{0}, a$, and $b$.
In this procedure the polarity $A$ is not involved, because it is regarded as fixed parameter.
The flux determination done under both $A^{+}/A^{-}$ hypotheses gives the two $J^{\pm}$ fluxes of Eq.(\ref{weightedflux}).
The best model is shown in Fig.\,\ref{interpol} as thick long-dashed line, along with 32 flux calculations of all adjacent grid nodes.
The model is superimposed to the data from PAMELA corresponding to December 2011 (BR 2445).
During this epoch the HMF was in well-defined negative polarity state. All fluxes in the figure are calculated for $A=-1$, \ie, with $\mathcal{P}=1$.
%
\begin{figure}[htb]
\centering
\includegraphics[width=0.40\textwidth,scale=0.30]{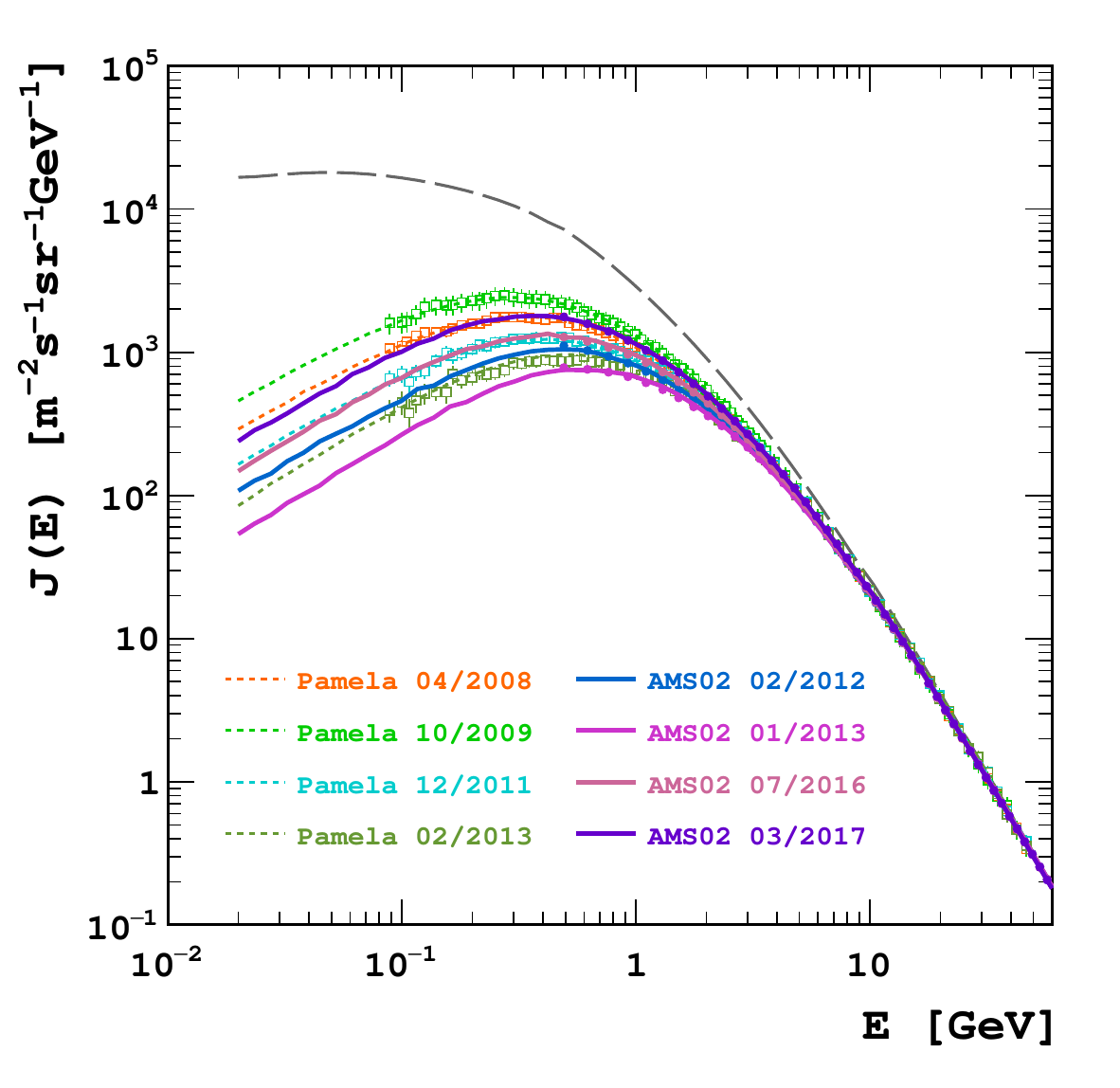}
\caption{Best-fit fluxes for selected data sets corresponding to PAMELA (dotted lines) and AMS-02 (solid lines) measurements.
  The long-dashed line represents the proton LIS used in this work.}
\label{bestflux}       
\end{figure}

\section{Results and discussion}    
\label{Sec::Results}                

Here we present the results of the fitting procedure described in Sect.\,\ref{Sec::TheInference} and implemented using the considered
data set on \GCR protons of Sect.\,\ref{Sec::TheData}.
We found that the agreement between best-fit model and the measurements on the fluxes of \GCR protons was in general very good for all
the data sets and over the whole rigidity range.
In Fig.\,\ref{bestflux} the best-fit models for the proton fluxes are shown as colored lines for some selected epochs, along with the \GCR proton LIS.
The calculations are compared with the data from experiments PAMELA and AMS-02 at the corresponding epochs.
The long-dashed line represents the proton LIS model used in this work and presented in Sect.\,\ref{Sec::ProtonLIS}.
%
\begin{figure*}[hbt]
\centering
\includegraphics[width=0.90\textwidth,scale=0.50]{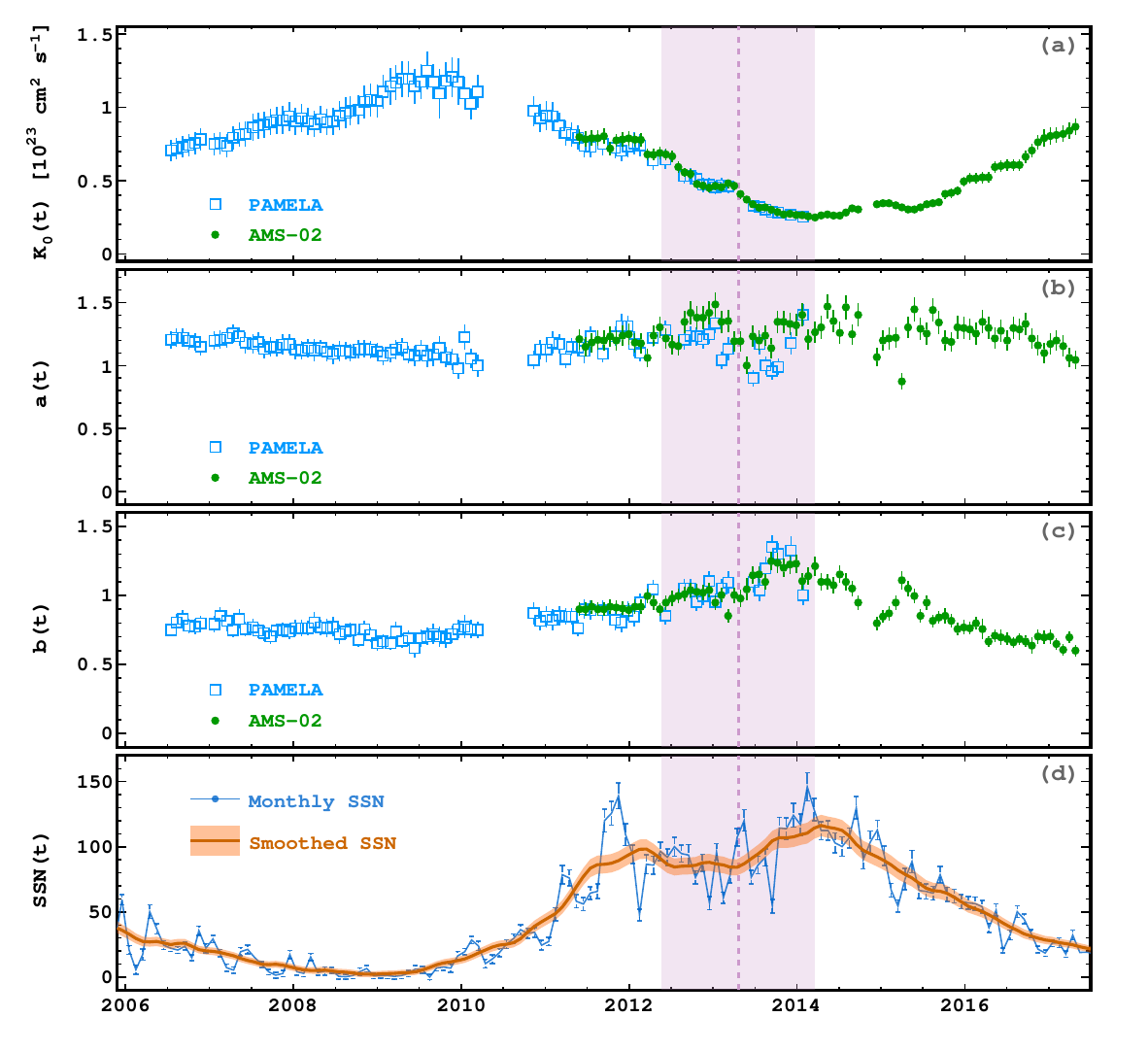} 
\caption{Results for the best-fit model parameters $K_{0}$, $a$, and $b$ determined using the time-resolved proton flux measurements
  from PAMELA (open squared) and AMS-02 (filled circles). 
  In panel (d), the monthly averaged and smoothed SSN is shown. The vertical dashed line indicates the reversal epoch $T_{\rm{rev}}$
  and the shaded area around it shows the transition epoch where the HMF polarity is weakly defined.
  }
\label{Fig::TransportParametersVSTime}
\end{figure*}

\subsection{Temporal dependencies} 
\label{Sec::TemporalDependencies}  

The main results on the parameter determination procedure are illustrated in Fig.\,\ref{Fig::TransportParametersVSTime}.
The figure shows the best-fit model parameters $K_0$, $a$, and $b$ as
function of the epoch corresponding to the measurements of AMS-02 (filled circles) and PAMELA (open squares).
The vertical dashed line and the shaded area around it represent the reversal phase, as in the previous figures. 
As a proxy for solar activity, Fig.\,\ref{Fig::TransportParametersVSTime}d shows the monthly SSN data.
The solid line shows the smoothed SSN values, obtained with a moving average within a time window of 13 months, along with its uncertainty band.
It can be seen that the diffusion parameters show a remarkable temporal dependence, and such a dependence is well correlated with solar activity.
From the figure, it can be seen that the normalization of the parallel diffusion coefficient $K_{0}$ 
shows a clear temporal dependence.
The diffusion normalization appears to be maximum in the $A<0$ epoch before reversal ($t\ll{T_{\rm{rev}}}$), and in particular during the
unusually long solar minimum of 2009-2010.
The minimum of $K_{0}$ is reached during solar maximum in 2014, about one year after polarity reversal.
From the comparison between panel (a) and panel (d),
the $K_{0}$ parameter appears anti-correlated with the monthly SSN.
Physically, larger values of $K_{0}$ imply faster \GCR diffusion inside the heliosphere, thereby causing a milder attenuation of the LIS, \ie,
giving a higher flux of cosmic protons in the GeV energy region. In contrast, lower $K_{0}$ values imply slower \GCR diffusion which is typical
in epochs of high solar activity where the modulation effect is significant.
Qualitatively, this behavior can be interpreted within the Force-Field approximation where, in fact, positive correlation is expected
between SSN and the modulation potential $\phi\propto 1/K_{0}$ \citep{Tomassetti2017BCUnc}.
Within the framework of the Force-Field model, the parameter $\phi$
is interpreted as the average kinetic energy loss of \GCR protons inside the heliosphere.
For similar reasons, a positive correlation between the best-fit $K_{0}$-value and the \GCR flux intensity $J_{0}$ at a given energy
as can be noticed, in particular, from the comparison of Fig.\,\ref{Fig::TransportParametersVSTime}a with Fig.\,\ref{Fig::ReferenceFlux}.
Our finding are in agreement with earlier works \citep{Manuel2014,Tomassetti2017TimeLag,Corti2019}. 
During the reversal phase, the temporal evolution of the model parameters in Fig.\,\ref{Fig::TransportParametersVSTime}
is obtained using the weighted linear combination of model fluxes with opposite polarities given by Eq.(\ref{weightedflux}).
During this epoch, the diffusion of \GCRs is slow and the tilt angle $\alpha$ reaches large values, typically higher than $65^{\circ}$.

The inferred $K_{0}$-values and their temporal evolution are related to the level of magnetic turbulence in the heliospheric plasma.
As clear from the figure, the diffusion is faster when the Sun is quiet with low turbulence levels and vice-versa.
From Eq.(\ref{Eq::KvsR}), the \GCR diffusion coefficients are linked to the HMF intensity and its temporal evolution which, however,
from Fig.\,\ref{Fig::SolarParameters}, appears to be quite shallow in the epoch considered.
As recently suggested in \citem{Wang2019}, the relation between the diffusion coefficient and the magnitude of the local HMF
can be described by a power-law, but the two quantities obey to different relationships for ascending and descending phases of the Solar Cycle.
Physical explanation for these behaviors may involve temporal variations in the spectrum of heliospheric turbulence during the solar cycle \citep{Vaisanen2019,Zhao2018},
that we discuss in the following. Investigations on the correlations between solar and diffusion parameters are made in Sect.\,\ref{Sec::CrossCorrelations}.

\subsection{The evolving turbulence} 
\label{Sec::EvolvingTurbulence}      
%
The $a$ and $b$ parameters shown in Fig.\,\ref{Fig::TransportParametersVSTime} describe the rigidity dependence of \GCR diffusion
tensor $K_{\parallel}$ below and above the break value $R_{k}$.
These parameter can test how the Sun variability affects the spectrum of magnetic irregularities
of the heliospheric plasma, that is, its turbulence spectrum.
From figure, it can be noted that both parameters show a characteristic temporal dependence in the epoch considered.
In the negative polarity epoch of $t{\ll}T_{\rm{rev}}$, and in particular during solar activity minimum,
the spectral indices of \GCR diffusion are seen to vary smoothly and slowly with time. 

The two spectral indices show a different temporal dependence.
The index $a$ is found to be essentially time independent, with an average value of $a$ = 1.21$\pm$0.06, while
the index $b$ shows a distinct long-term evolution in the considered period.
During the long unusual minimum from 2006 to 2009, $b$ remains constant at a value of $b$ = 0.74$\pm$0.03,
as long as the solar activity is quiet and the corresponding number of monthly sunspots is below $\sim$\,50. 
Subsequently, in $\sim$\,2010-2011, when the ascending phase of the solar cycle sets in, $b$ starts to increase steadily.
During this period, the \GCR flux decreases steadily as well.
The increase keeps going during the whole reversal phase,  \ie, at full maximum solar activity.
Here the $b$ parameter reaches an average maximum value of  1.3\,$\pm$\,0.07.
After this phase and during the flux recovery phase in the positive polarity epoch,  the index $b$ decreases steadily during the descending phase of the solar cycle,
until it recovers the values of the previous solar minimum.
Instead, the index $a$ shows no prominent features over the whole descending phase.

It should be noted, however, that the $a$ parameter is poorly constrained in the $A>0$ phase, because the
AMS-02 data are available only above 1\,GV of rigidity, and thus they are not highly sensitive to this parameter.
From the figure, it can be seen that the index $b$ is negatively correlated with the diffusion normalization parameter $K_{0}$:
during minimum, where $K_{0}$ is large and the \GCR diffusion is therefore fast, its rigidity dependence is shallow ($b\approx$\,0.8) in comparison to
solar maximum, where diffusion is slow and its rigidity dependence is more pronounced ($b\approx$\,1.3).  
Since the two indices are related to the power spectrum of the heliospheric turbulence, they could be used to infer the
spectral index $\nu$ of the power spectrum density of HMF irregularities (see Sect.\,\ref{Sec::Transport}).
Keeping in mind that $\lambda_{\parallel}\propto R^{2-\nu}$, the index $a$ is related to the power spectrum density
in the energy-containing range, while the index $b$ is related to the power spectrum
in the inertial range of the turbulent energy cascade of HMF.
The results indicate that the diffusion spectrum in the energy-containing regime does not depend on the solar activity,
while, in the inertial range, the spectrum appears to evolve as a function of the solar activity, with a clear delayed peak at the solar maximum.
The spectral index of the turbulence in the energy-containing range is $\nu_{ec}$ = 0.79$\pm$0.13 over all the period examined in this work,
while in the inertial range the spectral index evolves from $\nu_{in}$ = 0.74$\pm$0.08 at solar minimum to $\approx$1.3$\pm$0.15 during the solar maximum.

The temporal and rigidity dependence of the \GCR mean free path $\lambda_{\parallel}(t,R)$ can be determined
from Eq.(\ref{Eq::KvsR}) using our best-fit parameters. At the $R{\approx}1$\,GV rigidity scale,
our $\lambda_{\parallel}$ is found to range between 0.05 AU and 0.3 AU, depending on solar activity.
This result is in excellent agreement with
the large collection made in \citem{Palmer1982} of observational measurements on the scattering mean free path \citep{Tautz2013}.
In addition, our result show that the \GCR variability involves the rigidity dependence of the diffusion tensor, in particular via the spectral indices $a=a(t)$ and $b=b(t)$. 
An important implication of this finding is that the parallel diffusion coefficient cannot be write as a product $K_{\parallel}(t,R)= f(t){\times}g(R)$,
where a universal rigidity dependence $g(R)$ is modulated in amplitude by means of a factorized function $f(t)$ \citep{Tomassetti2018PHeVSTime,Manuel2014}.
Mathematically, this makes the $K_{\parallel}(t,R)$ function of Eq.(\ref{Eq::KvsR}) a \emph{non separable} function of rigidity and time variables.
Physically, it indicates that the HMF turbulence spectrum varies significantly over the solar cycle, depending on the cycle phase.
In particular, the power spectrum is observed to be steeper around solar maximum and flatter during solar minimum, with a quasi-periodical pattern.
The temporal variability of HMF turbulence is also studied from the analysis of neutron monitor data \citep{Vaisanen2019}.
These findings suggest that during epochs of quiet activity, kinetic self-organized turbulence dominates the \GCR spectrum, such as, \eg, a Kolmogorov-type cascade,
while random processes and transient events in the heliosphere play a key role during high-activity epochs of the solar cycle.
The use of wider sets of data may allow to provide better clarification on such a behavior.

\subsection{Cross-correlations} 
\label{Sec::CrossCorrelations}  

\begin{figure}[t]
\centering
\includegraphics[width=0.40\textwidth,scale=0.5]{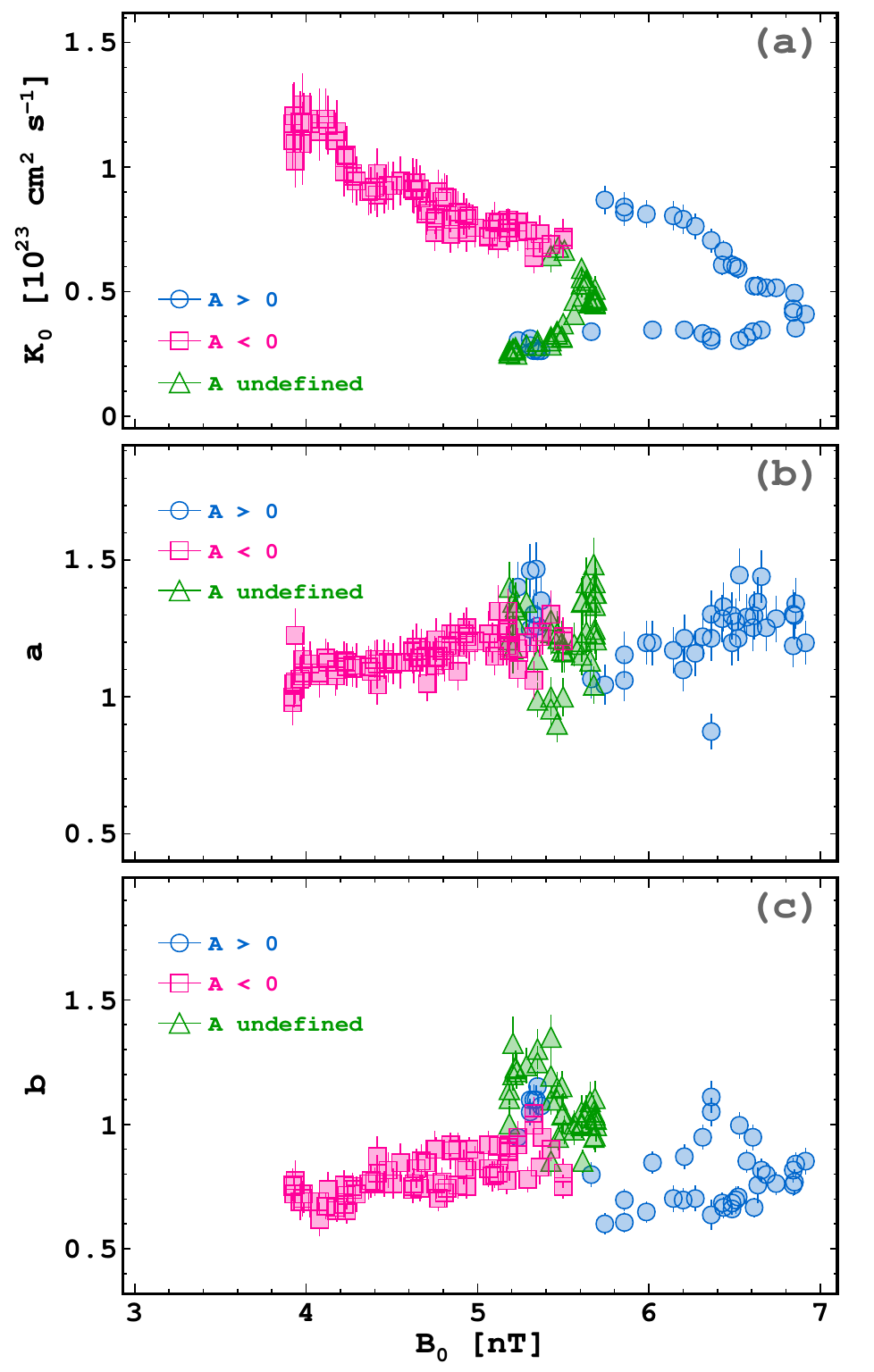} 
\caption{Scatter plots showing the correlation between the best-fit diffusion parameters and BMA reconstruction of the local HMF $\widehat{B}_{0}$.
  The results are divided in groups of positive polarity (blue circles), negative polarity (pink squares), and reversal phase (green triangles).
}
\label{Fig::K0vsB0} 
\end{figure}
%
%
We now inspect the running cross-correlation between solar and transport parameters.
Figure\,\ref{Fig::K0vsB0} displays the scatter diagrams of the best-fit diffusion parameters against the BMA reconstruction of the local
HMF value, $\widehat{B}_{0}$ (left column) and the HCS tilt angle $\alpha$ (right column).
In panel (a), the diffusion normalization parameter $K_{0}$ is shown. The different markers are used to indicate the reconstructions
obtained during epochs of positive (blue circles) and negative polarity (pink squares), as well as during reversal phase (green triangles).
This behavior can be compared with the one found by \citet{Wang2019} where, from an analysis of the ascending and descending phases of the
solar cycle (both during negative polarity) two distinct power-law relations were observed between diffusion coefficient and local HMF magnitude.
Our results confirm the relationship between $K_{0}$ and $B_{0}$
becomes complex when the examination is done over a large fraction of the solar cycle that include polarity changes.
In particular, two distinct relationships can be observed for $A<0$ and $A>0$ polarity conditions.
%
%
\begin{figure}[t]
\centering
\includegraphics[width=0.40\textwidth,scale=0.5]{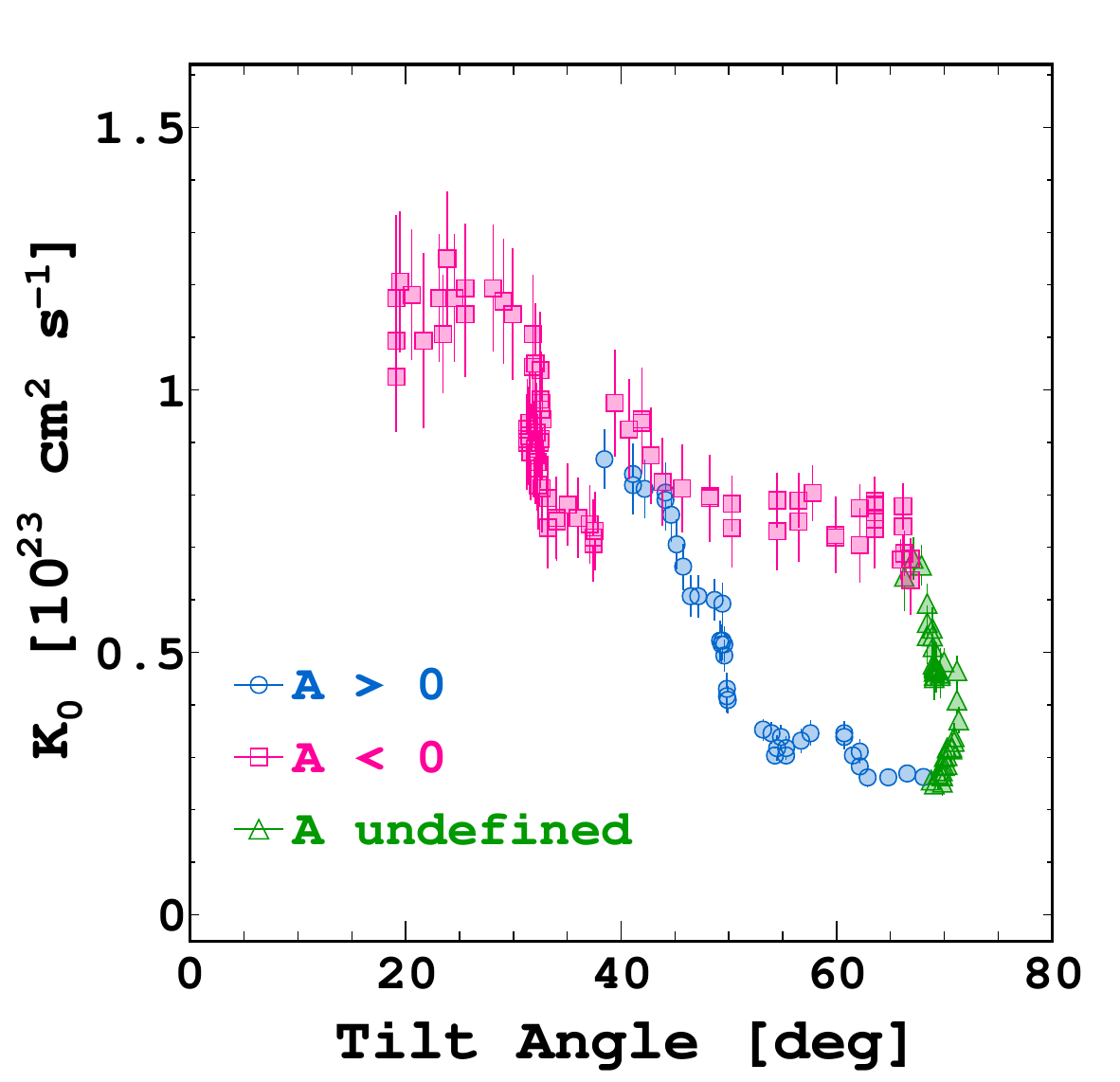}
\caption{Scatter plot of the best-fit parameter $K_{0}$ versus the HCS tilt angle.
  The results are divided in groups of positive polarity (blue circles), negative polarity (pink squares),
  and reversal phase (green triangles).}
\label{Fig::K0vsTA}
\end{figure}
%
%
%
%
Regarding the correlation between the spectral index parameters $a$ and $b$ with the HMF magnitude $\widehat{B}_{0}$, smoother relationships were found.
The index $a$ is nearly constant with time,
while the index $b$ increases slowly during solar maximum, \ie, during the reversal phase.
Both parameters are seen to depend only weakly on the polarity phase, and no particular cross-correlation is observed between two spectral indices.
The scatter plot of $K_{0}$ versus tilt angle is also shown, in Fig.\,\ref{Fig::K0vsTA} where, again,
the different style of the markers refer to the different phases of solar activity.
The dependence is similar to that observed with the HMF intensity, showing a pronounced negative correlation and a characteristic modulation loop.

%
\begin{figure}[hbt!]
\centering
\includegraphics[width=0.40\textwidth,scale=0.5]{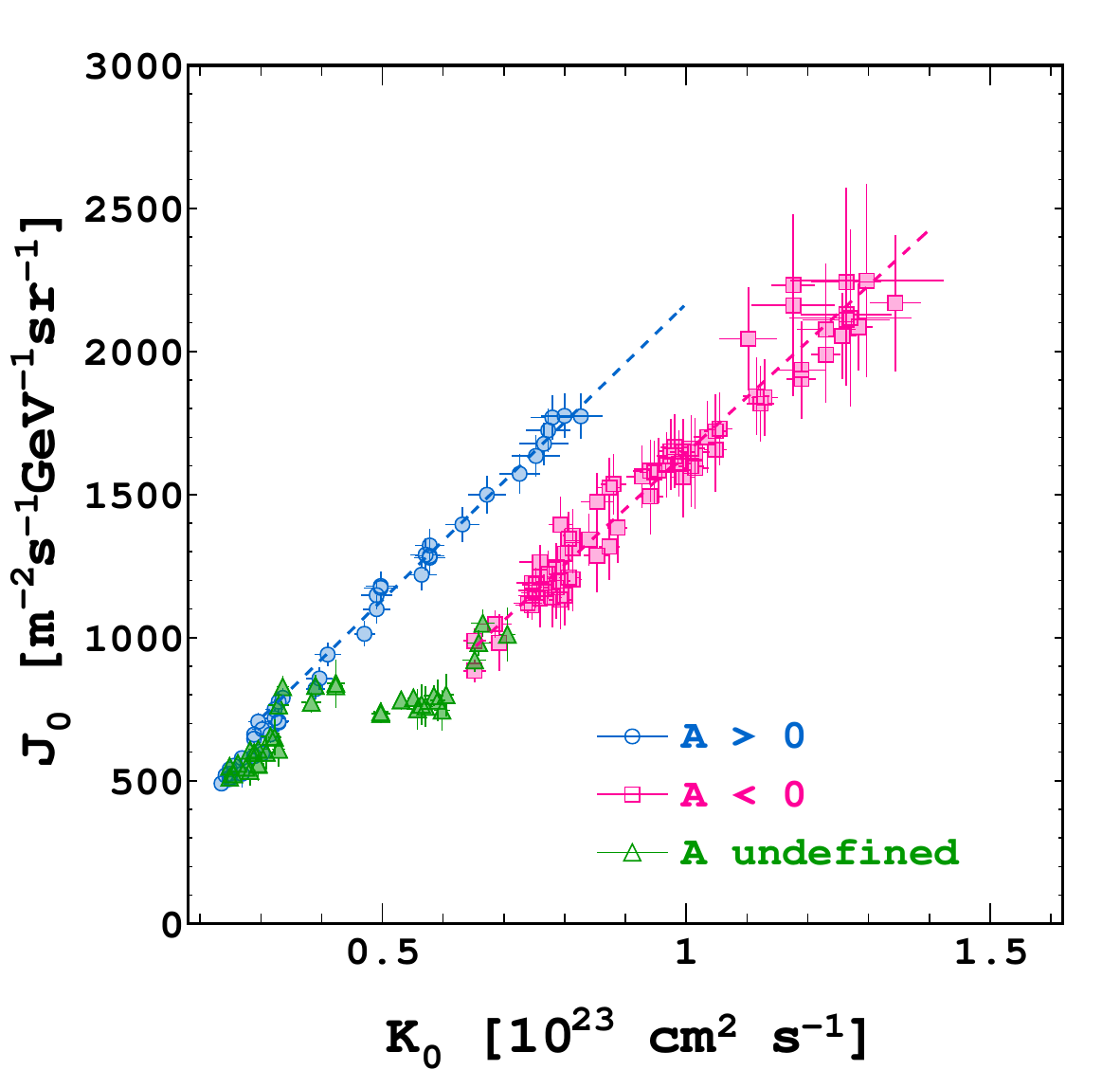} 
\caption{Scatter plot of the \GCR flux $J_{0}$, evaluated at the reference energy interval $E_{0} = 0.49-0.62$\,GeV,
  against the normalization factor of the diffusion tensor $K_{0}$.
  The color coding is the same of Fig.\,\ref{Fig::K0vsTA}.
  During the phases of well defined polarity, $J_{0}$ shows a distinct dependence on the diffusion
  strength parameter that has been fit with Eq.(\ref{jref_fit}) (dashed lines).}
\label{Fig::CorrelationJ0vsK0}     
\end{figure}
%
%
The correlation between the  flux intensity $J_{0}$ and the diffusion normalization $K_{0}$ is shown in Fig.\,\ref{Fig::CorrelationJ0vsK0}.
In this figure, the flux intensity $J_{0}$ is extracted from the data at the reference kinetic energy $E_{0}=0.49-0.62$\,GeV, as in Fig.\,\ref{Fig::ReferenceFlux}, 
while $K_{0}$ is the best-fit value at the corresponding epoch.
From the figure, the \GCR flux intensity appears in general well correlated to the normalization factor of the diffusion coefficient,
which appears to be the driving parameter of the modulation model.
It can also be seen that relationship between $J_{0}$ and $K_{0}$ is remarkably linear during epochs of well-defined polarity.
We describe it with the following empirical relation:
\begin{equation}
\label{jref_fit}
J_{0}(K_{0}) = \eta K_{0} + J_{\rm{off}}  \,.
\end{equation}
By making separate fits for the two polarity epochs,
we obtained $\eta^{+} = (2212\pm 250)\times 10^{-23}$ for $A>0$, and $\eta^{-}=(1929\pm 260)\times 10^{-23}$\,$\rm{cm^{-4}GeV^{-1}sr^{-1}}$ for $A<0$.
The best-fit offset are $J_{\rm{off}}^{+} = -46\pm21$ for positive polarity, and $J_{\rm{off}}^{-} = -286\pm68$ $\rm{m^{-2}s^{-1}GeV^{-1}sr^{-1}}$ for negative polarity. 
The two fits are shown in Fig.\,\ref{Fig::CorrelationJ0vsK0} as dashed line.
It is interesting to note that, within the fitting errors, the two slopes $\eta^{+}$ and $\eta^{-}$ turned out to be consistent each other, \ie,
the slope of  $J_{0}(K_{0})$ is polarity and charge-sign independent.
Polarity-effect results into different offsets $J_{off}^{\pm}$ for the two phases.
This result may help to quantify the effects of drift motions to the \GCR modulation.
The diffusion coefficient appears to be independent upon the $\hat{q}A$ sign product, as indicated by the consistency between
$\eta^+$ and $\eta^-$ values from the fit.
For a given $K_{0}$ value, the resulting difference in the fluxes is only due to the opposite directions of the net drift and convective flux
for epochs of opposite polarities. 
The quantity $\Delta J \equiv J_{\rm{off}}^{+} - J_{\rm{off}}^{-}$ can be used as a measurement of the net effect of drift
on the total \GCR flux, for a given level of \GCR diffusion.

\begin{figure*}[th!]
\centering
\includegraphics[width=0.90\textwidth,scale=0.50]{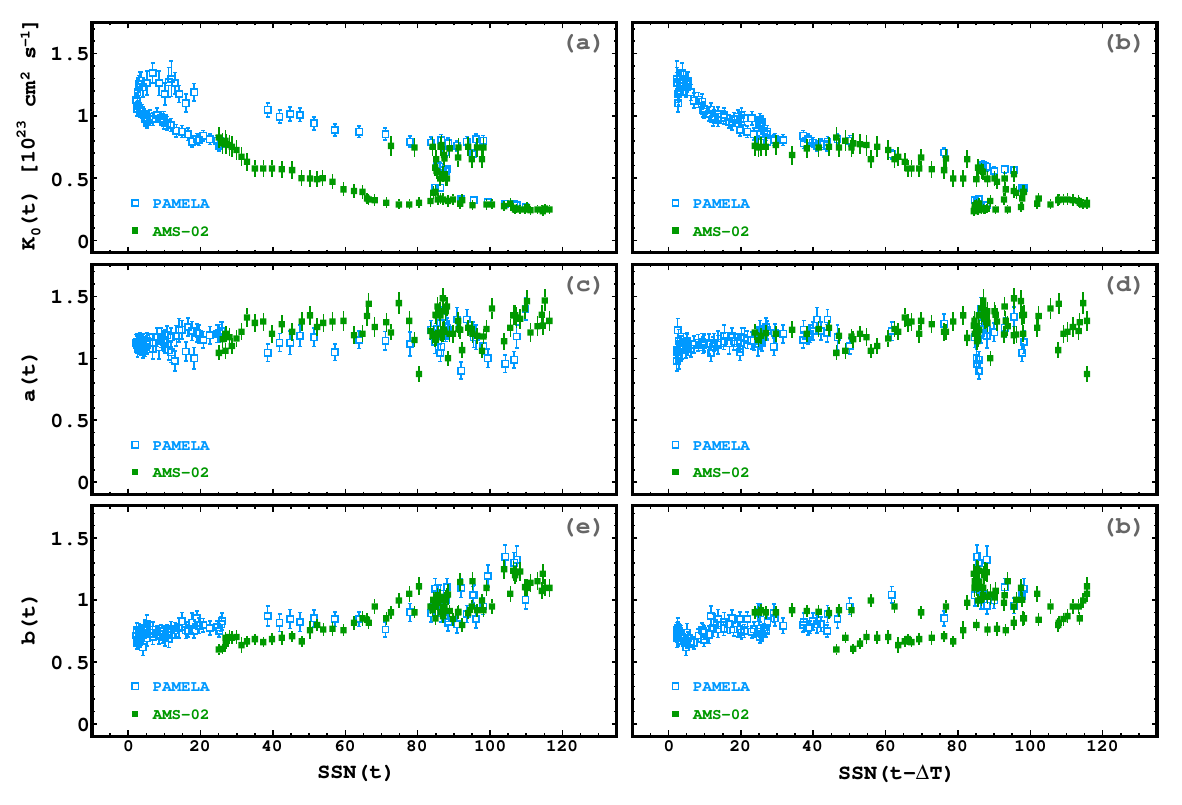} 
\caption{Model parameters as a function of the SSN. The left column displays the parameters as a function of $SSN(t)$ versus the
  SSN at the same same epoch, the right column displays the parameters as a function of $SSN(t-\Delta T_{\rm{lag}})$.
  The AMS-02 and PAMELA data are represented by the red and black dots, respectively.}
\label{Fig::ParametersSSNCorrelations}
\end{figure*}
%
%
We also note that in the figure, the fit results obtained under periods of undefined polarity (green triangles)
connect smoothly the two regimes. In this epoch the role of drift is not well understood,
but the flux $J_{0}$ remains correlated with $K_{0}$.
To close the loop, it may take an entire cycle of magnetic polarity.

\subsection{Lags and loops} 
\label{Sec::LagsAndLoops}   
%
From Fig.\,\ref{Fig::TransportParametersVSTime}, it can be noticed that a time shift of a few month
is present between the smoothed SSN (the ${S}(t)$ function) and the best-fit modulation parameters $K_{0}(t)$, $a(t)$ and $b(t)$.
For instance, the highest \GCR flux intensity was reached around October 2009, with $J^{\rm{max}}=$\,2289$\pm$220 $\rm{m^{-2}s^{-1}GeV^{-1}sr^{-1}}$,
\ie, about eight months after the SSN minimum of February 2009.
Similarly, the minimum flux intensity was observed around February 2014, $J^{\rm{min}}=$\,498$\pm$\,23 $\rm{m^{-2}s^{-1}GeV^{-1}sr^{-1}}$,
while solar maximum occurred in April 2013.
To estimate the average time lag between $K_{0}(t)$ and the smoothed SSN ${S}(t)$,
we compare the correlation between $K_{0}(t)$ and ${S}(t - \Delta T_{\rm{lag}})$. 
The best-value for the lag $\Delta{T_{\rm{lag}}}$ can be obtained by a scan of $\Delta T_{\rm{lag}}$,
in order to determine the Pearson linear correlation coefficient $\rho$ as function of $\Delta T_{\rm{lag}}$.
The $\Delta{T_{\rm{lag}}}$ parameter which maximizes $\rho$ is then taken as best estimate of the average time lag between the SSN and \GCR modulation parameters.
For the analyzed period, we obtain $\Delta T_{\rm{lag}}\,= 11.4\pm 1.4$ months.
Thus, on average, the modulation of \GCRs observed at the epoch $t$ is related to manifestations of solar activity at the epoch $t - \Delta T_{\rm{lag}}$. 
The correlation between diffusion parameters and smoothed SSN is shown in Fig.\,\ref{Fig::ParametersSSNCorrelations},
where the model parameters at the epoch $t$ are shown as a function of the SSN at the same epoch (left column) and at the
epoch $t - \Delta T_{\rm{lag}}$ (right column).
In general, when the time lag is not taken into account, the diffusion normalization $K_{0}(t)$ appears as a multivalued function of SSN,
showing a characteristic hysteresis structure over the different phases of the solar cycle.
When the lag is taken into account, the curve of $K_{0}$ vs SSN shrinks, approaching a single-valued function. This would allow, in principle,
to forecast the modulation parameters at the epoch $t$ from observations of SSN made in advance by $\Delta T_{\rm{lag}})$.
However, the $a$ and $b$ parameters versus the delayed SSN do not show clear one-to-one relationships,
which suggests that the use of a single lag value may be a too simplistic approach.
The calculated lag depends weakly on the BMA averages used to define the heliosphere status. On the other hand,
the BMA procedure of Sect.\,\ref{Sec::TheParameters} is well motivated by the observation of such a lag.
In this respect, an estimate of the uncertainty on ${\Delta}T_{\rm{lag}}$ can be done by varying
the time window $T_{\rm{BMA}}$ used to get the average conditions ($B_{0}$ and $\alpha$) of the heliosphere. 
Our estimation of ${\Delta}T_{\rm{lag}}$ is fairly consistent with other recent works \citep{Tomassetti2017TimeLag,RossChaplin2019,ChowdhuryKudela2018}.
Nonetheless, there are some discrepancies with the reported values if one account for even/odd cycle dependence of the lag.
Our estimation of the time lag lies in solar cycle 24, but it appears longer than that reported in previous even-numbered solar cycles,
though it is comparable to the lag observed in odd-numbered solar cycles \citep{Aslam2015,Singh2008,Iskra2019}. 
In this respect, as well as in other characteristics, cycle 24 is unusual when compared to previous even cycles.
Other differences may be related to the rigidity of \GCR particles, as past studies are based on neutron monitors rates.
The global dependence of the time lag upon the solar cycle and on the rigidity of the \GCR particles will be addressed in a forthcoming paper.

\section{Conclusions and discussion} 
\label{Sec::Conclusions}             

Thanks to the recent availability of time-resolved data from space, the study of \GCRs in the heliosphere has become an active topic of investigation.
In particular, the recent data released by AMS-02 and PAMELA on the monthly evolution of proton and helium  permits new investigation of the
solar modulation phenomenon over a large fraction the of solar cycle.
These data have triggered new efforts at establishing advanced models of \GCR propagation in heliosphere \citep{Luo2019,Boschini2018,Boschini2020,Ngobeni2020,Bobik2021}.
In particular, many recent studies were focused on specific aspects of the \GCR modulation such as, \eg,
the particle dependence of \GCR diffusion \citep{Tomassetti2018PHeVSTime,Corti2019},
the relationship between modulation and solar activity proxies \cite{Wang2019,Wang2020},
the derivation of improved LIS evaluation \citep{Boschini2020,Zhu2018},
or the extraction of \GCR modulation parameters using statistical inference \citep{Corti2019},
which is also the main goal of the present paper.
More specifically, in this paper, we have investigated the propagation of Galactic \GCRs in the heliosphere using 
a numerical model based on stochastic simulations and calibrated by means of a large set of experimental data.
The data consist of time-series of \GCR proton fluxes reported by AMS-02 and PAMELA experiments in low Earth orbit.
The measurements are made on 27-day basis, corresponding to a solar rotational period,
and cover a time range of 11 years, corresponding to a solar cycle period. 
The sample include epochs of very different solar conditions such as solar minimum, solar maximum,
ascending and descending phases, as well as positive and negative HMF polarity states.
The time range and resolution of these data is therefore optimal for the study long-term modulation of Galactic \GCRs,
and in particular, for investigating influence of solar variability in the diffusive propagation of \GCRs in the heliospheric turbulence. 

In our calculations we have used, as time-dependent physical inputs, BMA values of the tilt angles $\alpha$ of the HCS,
the local HMF strength at 1 AU $B_{0}$, and the magnetic polarity $A$. These quantities constitute a very good proxies for solar activity.
In this analysis, we have been focused on the parameters describing the temporal and rigidity dependence of \GCR diffusion.
We have determined the time-series of the diffusion normalization, $K_{0}$, and that of the spectral indices $a$ and $b$ that control the dependence of \GCR diffusion upon rigidity.

In practice, to perform a statistical inference using the data, and to account for the evolving conditions of the heliospheric plasma, 
we have built a large array of differential energy fluxes $J(E)$, evaluated at Earth's location, corresponding to 938,400 parameter configurations. 
To sample such a 6-dimensional parameter space, we have simulated about 14 billions trajectories of cosmic protons in the interplanetary space.
Each simulated particles was backwardly propagated from Earth's vicinity to the heliospheric boundaries.
The array of models generated in this work can be used to estimate the modulation parameters of \GCR protons at any epoch and
for any set of experimental data, ranging from 20\,MeV to hundreds GeV of kinetic energy.
We also note that in our model, the time dependence of the problem is treated by providing a time series
of steady-state solutions for $J_{\rm p}$ associated with a time series of input parameters $k_{0}$, which is a simplification.
Such an approach stands as long as the timescales between \GCR transport in the heliosphere does not exceed the
analyzed changes in solar activity. To extend the analysis to smaller time-scale (\eg, daily) or to lower energies (\eg, MeV-scale),
a time-dependent solution of the Parker's equation should be considered.
Nonetheless, we also stress that the time-series of best-fit parameters derived in this work should be regarded as effective values,
averaged over the \GCR propagation histories, not necessarily representing the instantaneous conditions of the heliospheric plasma.

Our approach is also simplified in several aspects, for example regarding the
rigidity and spatial dependence of the diffusion tensor, or its perpendicular components.
Nonetheless, in comparison to our earlier works, we have introduced several new recipes 
that capture most of the relevant features of \GCR propagation in the heliosphere.
The agreement of our calculations with the \GCR flux data is very satisfactory.
As we have shown, using \GCR proton data, it is possible to determine the detailed evolution of the
rigidity dependence of the diffusion coefficient with the solar activity, ad thus,
the physical nature of the turbulence embedded in the frozen-in HMF carried out by the SW.
Our findings indicate that solar variability has an important effect on the turbulence spectrum of HMF irregularities,
and an imprint of this mechanism can be observed in the rigidity dependence of the diffusion tensor.
In particular, we have reported a remarkable long-term dependence for the two spectral indices $a$ and $b$. 
These results show that the turbulence regime evolves with time, following the solar cycle, and thus
the temporal and rigidity dependencies of \GCR diffusion coefficients cannot be described by a 
separable function of the type $K_\parallel(t,R){\equiv}K_{0}(t){\times}f(R)$. 
In this respect, we remark that the time-rigidity separability for \GCR diffusion is assumed by several models of solar modulation,
although such an assumption is not supported by theoretical considerations \citep{Moraal2013,Manuel2014,Wang2019}. 
Moreover, the study of the correlation between solar and diffusion parameters reveals charge-sign dependent
features in the \GCR modulation effect, such as different patterns for the different phase of the HFM polarity cycle.

We remark that solar cycle 24 has been unusual when compared to the previous cycles, therefore also the \GCR modulation conditions were unusual.
The solar minimum between cycles 23 and 24 was quite longer and deeper than expected \citep{Potgieter2014,Aslam2015}.
while the maximum of cycle 24 was the smallest recorded in a century of standardized SSN observations, and with a double-peak structure \cite{CletteLefevre2016}.
In our analyzed data sample, the correlation between \GCR flux modulation and solar activity as measured by the SSN is apparent.
The \GCR proton intensity modulation, in anti-phase with solar activity, in the considered period shows an average time lag of about 11 months.
A next phase of this work is to study the dependence of the lag on solar activity parameters (such as SW speed or HMF polarity)
and \GCR transport properties (such as diffusion or drift coefficients), in order to understand the dynamics of the physical mechanisms
behind the solar modulation phenomenon.
Further steps also include the implementation of a better description of the HMF, of the diffusion tensor and
the drift reduction factor during solar maximum.
In particular, we assumed ``full drift'' at any phase of the cycle, including the HMF reversal epoch
where the modulated flux of \GCRs was modeled as superposition of fluxes with positive and negative polarity states.
While our approach provided a good description of the flux evolution in the reversal region,
one may argue that large-scale drift may be suppressed during solar maximum  due to the more chaotic structure of the HMF. 
This idea can in principle be tested using the data.  In particular, the availability of time-dependent
measurements on \GCR antiprotons will be precious to study the modulation effect across solar maximum.
Data of the temporal dependence of \GCR antiprotons are still lacking, but the AMS-02 experiment has the capability of making such a measurement.

\section*{Acknowledgement} 
%
We acknowledge the support of Italian Space Agency (ASI) under agreement ASI-UniPG 2019-2-HH.0. 
B.K. acknowledge support from agreement ASI-INFN 2014-037-R.1-2017, M.G. and F.D. from ASI-INFN 2019-19-HH.0. 
The cosmic ray data used in this work have been retrieved through the \emph{Cosmic Ray Data Base} of the ASI Space Science Data Center. 
Data on the Sun's polar magnetic field and tilt angle are taken from the Wilcox Solar Observatory at Stanford University.
Interplanetary HMF data of the Advanced Composition Explorer have been downloaded from the OMNIWeb service of the NASA Space Physics Data Facility.
Data on the sunspot numbers are provided by the SIDC-SILSO center at the Royal Observatory of Belgium, Brussels.


\end{document}